\newif\ifmakebbl
\documentclass[11pt]{article}
\usepackage[english]{babel}
\usepackage{epic,eepic}
\usepackage{epsfig}
\usepackage{latexsym}
\usepackage{times}
\usepackage{theorem}

\newcommand{\tuple}[1]{\langle#1\rangle} 

\newcommand{\efi}{\epsilon^{\gvp}_i}
\newcommand{\egi}{\epsilon^{\psi}_i}
\newcommand{\ev}{\epsilon(v)}
\newcommand{\evar}[1]{\epsilon(#1)}

\newcommand{\nbls}{\vspace*{-1\baselineskip}}
\newcommand{\hnbls}{\vspace*{-.5\baselineskip}}
\newcommand{\tnbls}{\vspace*{-.33\baselineskip}}

\newcommand{\qed}[0]{\hspace*{0mm}\hfill $\Box$\vspace{3mm}}
\newcommand{\ol}[1]{{\overline#1}}
\newcommand{\proof}{\noindent {\bf Proof}.\ }

\newcommand{\raf}[1]{(\ref{#1})}

\newcommand{\OR}{\bigvee}
\newcommand{\AN}{\bigwedge}

\newcommand{\hspacea}{\hspace*{0cm}}
\newcommand{\hspaceb}{\hspace*{.35cm}}
\newcommand{\hspacec}{\hspace*{.45cm}}
\newcommand{\hspaced}{\hspace*{1.15cm}}
\newcommand{\hspacee}{\hspace*{1.71cm}}
\newcommand{\hspacef}{\hspace*{2.01cm}}

\newcommand{\ga}{\alpha}
\newcommand{\gb}{\beta}
\newcommand{\gc}{\gamma}

\newcommand{\gD}{\Delta}
\newcommand{\gG}{{\it PI}}

\newcommand{\gvp}{\varphi}
\newcommand{\Tw}{{\it Tw}}

\newcommand{\NP}{{\rm NP}}
\newcommand{\coNP}{\textrm{co-NP}}

\newcommand{\Pol}{{\rm P}}
\newcommand{\nondet}[1]{{#1\textrm{-}\Pol}}
\newcommand{\betapol}[1]{{\beta_{#1}\Pol}}
\newcommand{\cobetapol}[1]{{{\rm co}\textrm{-}\beta_{#1}\Pol}}

\newcommand{\Tr}{\mathit{Tr}}

\newtheorem{theorem}{Theorem}[section]

\newtheorem{lemma}[theorem]{Lemma}
\newtheorem{proposition}[theorem]{Proposition}
\newtheorem{corollary}[theorem]{Corollary}

\newtheorem{remark}{Remark}[section]
\newtheorem{definition}{Definition}[section]

\global
\global

\newcommand{\nop}[1]{}

\evensidemargin=\oddsidemargin

\newcounter{myenumctr}

\newenvironment{myitemize}{\begin{list}{$\bullet$}{\setlength{\leftmargin}{0pt}
\setlength{\itemindent}{\labelwidth}}}
{\end{list}}

\title{
\vspace*{-2\baselineskip}
\bf New Results on Monotone Dualization and \\ Generating Hypergraph
Transversals%
\thanks{A shorter version of this paper 
appears in: Proceedings of the 34th ACM Symposium on Theory of Computing (STOC-02), May 19-21, 2002, Montreal, Quebec, Canada.}
\\~\\[-2ex]}

\author{Thomas Eiter\thanks{Institut f\"{u}r
Informationssysteme, Technische Universit\"{a}t Wien, Favoritenstra{\ss}e
9-11, A-1040 Vienna,
Austria. Email: eiter@kr.tuwien.ac.at
}
\qquad Georg Gottlob\thanks{Institut f\"{u}r
Informationssysteme, Technische Universit\"{a}t Wien, Favoritenstra{\ss}e
9-11, A-1040 Vienna,
Austria. Email: gottlob@dbai.tuwien.ac.at
}
\qquad
Kazuhisa Makino\thanks{Division of Systems Science,
Graduate School of Engineering Science,
Osaka University,
Toyonaka, Osaka, 560-8531, Japan. Email: makino@sys.es.osaka-u.ac.jp}
}

\date{}

\begin{document}

\maketitle


\begin{abstract}
We consider the problem of dualizing a monotone CNF (equivalently,
computing all minimal transversals of a hypergraph), whose associated
decision problem is a prominent open problem in \NP-completeness. We
present a number of new polynomial time resp.\ output-polynomial time
results for significant cases, which largely advance the tractability
frontier and improve on previous results. Furthermore, we show that
duality of two monotone CNFs can be disproved with limited
nondeterminism. More precisely, this is feasible in polynomial time
with $O(\chi(n)\cdot\log n)$ suitably guessed bits, where $\chi(n)$ is
given by $\chi(n)^{\chi(n)} = n$; note that $\chi(n) = o(\log
n)$. This result sheds new light on the complexity of this important
problem.
\end{abstract}

\nbls

{\small
\begin{tabbing}
\> combinatorial enumeration, treewidth, hypergraph acyclicity, limited
nondeterminism.
\end{tabbing}
}

\section{Introduction}

Recall that the prime CNF of a monotone Boolean function $f$ is the
unique formula $\gvp=\bigwedge_{c\in S} c$ in conjunctive normal form
where $S$ is the set of all prime implicates of $f$, i.e., minimal
clauses $c$ which are logical consequences of $f$.  In this paper, we
consider the following problem:

\begin{center}
\fbox{
\parbox{5.5in}{
\smallskip

\centerline{Problem {\sc Dualization}}

\medskip
\centerline{\begin{tabular}{rl}
Input:&The prime CNF $\gvp$ of a monotone Boolean function
$f=f(x_1,\ldots,x_m)$. \\
Output:&The prime CNF $\psi$ of its dual $f^d =
\ol{f}(\ol{x_1},\ldots,\ol{x_m})$.
\end{tabular}}
\smallskip
}}
\end{center}

It is well known
that {\sc Dualization} is equivalent to the {\sc Transversal
Computation} problem, which requests to compute the set of all minimal
transversals (i.e., minimal hitting sets) of a given hypergraph ${\cal
H}$, in other words, the {\em transversal hypergraph} $\Tr({\cal H})$ of $\cal H$.
Actually, these problems can be viewed as the same problem, if the
clauses in a monotone CNF $\gvp$ are identified with the sets of
variables they contain.
{\sc Dualization} is a search problem; the associated decision problem
{\sc Dual} is to decide whether two given monotone prime CNFs $\gvp$
and $\psi$ represent a pair $(f,g)$ of dual Boolean functions.
Analogously, the decision problem {\sc Trans-Hyp} associated with {\sc
Transversal Computation} is deciding, given hypergraphs $\cal H$ and
$\cal G$, whether $\cal G = \Tr({\cal H})$.

{\sc Dualization} and several problems which are like transversal computation known to be
computationally equivalent to problem {\sc Dualization} (see
\cite{eite-gott-95}) are of interest in various areas such as
database theory (e.g.\ \cite{mann-raih-86,thi-86}), machine learning
and data mining (e.g., \cite{BGKM2001,BGKM2002,domingo-etal-99,guno-etal-97}), game
theory (e.g.\
\cite{Gur75,Ram90,Read78}), artificial intelligence (e.g.,
\cite{gogi-etal-98,kavv-etal-93,khar-95,reit-87}), 
mathematical programming (e.g., \cite{BEGKM2001}), and distributed
systems (e.g., \cite{garc-barb-85,ibar-kame-93}) to mention a few.

While the output CNF $\psi$ can be exponential in the size of $\gvp$,
it is currently not known whether $\psi$ can be computed in {\em
output-polynomial} (or {\em polynomial total}) {\em time}, i.e., in
time polynomial in the combined size of $\gvp$ {\em and} $\psi$. Any
such algorithm for {\sc Dualization} (or for {\sc Transversal
Computation}) would significantly advance the state of the art of
several problems in the above application areas. Similarly, the
complexity of {\sc Dual} (equivalently, {\sc Trans-Hyp}) is open since
more than 20 years now (cf.\
\cite{bioc-ibar-95,eite-gott-95,john-91,john-etal-88,lawl-etal-80}).

Note that {\sc Dualization} is solvable in polynomial total time on a
class $\cal C$ of hypergraphs iff {\sc Dual} is in PTIME for all pairs
$({\cal H},{\cal G})$, where $\cal H\in {\cal C}$ \cite{bioc-ibar-95}.
{\sc Dual} is known to be in $\coNP$ and the best currently known
upper time-bound is quasi-polynomial time
~\cite{fred-khac-96,gaur-99,tamaki-00}.  
Determining
the complexities of {\sc Dualization} and {\sc Dual}, and of
equivalent problems such as the transversal problems, is a prominent
open problem. This is witnessed by the fact that these problems are
cited in a rapidly growing body of literature and have been referenced
in various survey papers and complexity theory retrospectives,
e.g.~\cite{john-91,lova-92,papa-97}.

Given the importance of monotone dualization and equivalent problems
for many application areas, and given the long standing failure to
settle the complexity of these problems, emphasis was put on finding
tractable cases of {\sc Dual} and corresponding polynomial total-time
cases of {\sc Dualization}.  In fact, several relevant tractable
classes were found by various authors; see e.g.\
\cite{boro-etal-00,BGH93,BHIK97,cram-87,domingo-etal-99,eite-92,eite-gott-95,gaur-krish-00,maki-ibar-97,MI98,mish-pitt-97,PS94} and references
therein. Moreover, classes of formulas were identified on which {\sc
Dualization} is not just polynomial total-time, but where the
conjuncts of the dual formula can be enumerated with {\em incremental
polynomial delay}, i.e., with delay polynomial in the size of the
input plus the size of all conjuncts so far computed, or even with
{\em polynomial delay}, i.e., with delay polynomial in the input size
only. On the other hand, there are also results which show that
certain well-known algorithms for {\sc Dualization} are not
polynomial-total time. For example, \cite{eite-gott-95,mish-pitt-97}
pointed out that a well-known sequential algorithm, in
which the clauses $c_i$ of a CNF $\gvp=c_1\land\cdots\land c_m$ are
processed in order $i=1,\ldots,m$, is not polynomial-total time in general. Most
recently, \cite{taka-02} showed that this holds even if an
optimal ordering of the clauses is assumed (i.e., they may be
arbitrarily arranged for free).

\medskip

\noindent{\bf Main Goal.} The main goal of this paper is to present
important new polynomial total time cases of {\sc Dualization} and,
correspondingly, PTIME solvable subclasses of {\sc Dual} which
significantly improve previously considered classes. Towards this aim,
we first present a new algorithm {\sc Dualize} and prove its
correctness. {\sc Dualize} can be regarded as a generalization of a
related algorithm proposed by Johnson, Yannakakis, and
Papadimitriou~\cite{john-etal-88}. As other dualization algorithms,
{\sc Dualize} reduces the original problem by self-reduction to
smaller instances. However, the subdivision into subproblems proceeds
according to a particular order which is induced by an arbitrary fixed
ordering of the variables. This, in turn, allows us to derive some
bounds on intermediate computation steps which imply that {\sc
Dualize}, when applied to a variety of input classes, outputs the
conjuncts of $\psi$ with polynomial delay or incremental polynomial
delay. In particular, we show positive results for the following input
classes:


\begin{myitemize}
\item[$\bullet$] {\bf Degenerate CNFs.} We generalize the notion of
$k$-degenerate graphs~\cite{toft-95} to hypergraphs and define {\em
$k$-degenerate monotone CNFs} resp.\ {\em hypergraphs}. We prove that
for any constant $k$, {\sc Dualize} works with polynomial delay on
$k$-degenerate CNFs. Moreover, it works in output-polynomial time on
$O(\log n)$-degenerate CNFs.


\item[$\bullet$] {\bf Read-$k$ CNFs.} A CNF is {\em read-$k$}, if each
variable appears at most $k$ times in it. We show that for read-$k$
CNFs, problem {\sc Dualization} is solvable with
polynomial delay, if $k$ is constant, and in total polynomial time, if
$k=O(\log(\|\gvp\|)$.
Our result for constant $k$ significantly improves upon the
previous best known algorithm \cite{domingo-etal-99}, which has a higher
complexity bound, is not polynomial delay, and outputs the clauses of $\psi$
in no specific order. The result for $k=O(\log \|\gvp\|)$  is a non-trivial
generalization of the result in \cite{domingo-etal-99}, which was
posed as an open problem \cite{domingo-97}.


\item[$\bullet$] {\bf Acyclic CNFs.} There are several notions of
hypergraph resp.\ monotone CNF acyclicity~\cite{fagi-83}, where the
most general and well-known is $\alpha$-acyclicity. As shown
in~\cite{eite-gott-95}, {\sc Dualization} is polynomial total time for
$\beta$-acyclic CNFs; $\beta$-acyclicity is the hereditary version of
$\alpha$-acyclicity and far less general. A similar result for
$\alpha$-acyclic prime CNFs was left open.  (For non-prime
$\alpha$-acyclic CNFs, this is trivially as hard as the general
case.)  In this paper, we give a positive answer and show that for
$\alpha$-acyclic (prime) $\gvp$, {\sc Dualization} is solvable with
polynomial delay.


\item[$\bullet$] {\bf Formulas of Bounded Treewidth.} The {\em
treewidth}~\cite{robe-seym-86}
of a graph expresses its degree of cyclicity. Treewidth is an
extremely general notion, and bounded treewidth generalizes almost all
other notions of near-acyclicity. Following~\cite{chekuri-rajaraman-95}, we
define the treewidth of a hypergraph resp.\ monotone CNF $\gvp$ as the treewidth of its
associated (bipartite) variable-clause incidence graph.  We show
that {\sc Dualization} is solvable with polynomial delay
(exponential in $k$) if the treewidth of $\gvp$
is bounded by a constant $k$, and in polynomial total time
if the treewidth is $O(\log\log \|\gvp\|)$.


\item[$\bullet$] {\bf Recursive Applications of {\sc Dualize} and $k$-CNFs.}
We show that
if {\sc Dualize} is applied recursively and the recursion depth is
bounded by a constant, then {\sc Dualization} is solved in polynomial
total time.  We apply this to provide a simpler proof of the known
result~\cite{BGH93,eite-gott-95} that monotone $k$-CNFs (where each conjunct
contains at most $k$ variables) can be dualized in output-polynomial
time.
\end{myitemize}

After deriving the above results, we turn our attention (in
Section~\ref{sec:nondet}) to the  fundamental computational nature of
problems {\sc Dual} and {\sc Trans-Hyp} in terms of complexity theory.

\medskip

\noindent{\bf Limited nondeterminism.} In a landmark
paper, Fredman and Khachiyan~\cite{fred-khac-96} 
proved that problem {\sc Dual} can be solved in quasi-polynomial
time. More precisely, they first gave an algorithm~A solving the
problem in $n^{O(\log^2 n)}$ time, and then a more complicated
algorithm~B whose runtime is bounded by $n^{4\chi(n)+O(1)}$ where $\chi(n)$
is defined by $\chi(n)^{\chi(n)}=n$. As noted in~\cite{fred-khac-96},
$\chi(n)\sim \log n/\log\log n = o(\log n)$; therefore, duality
checking is feasible in $n^{o(\log n)}$ time. This is the best upper
bound for problem {\sc Dual} so far, and shows that the problem is
most likely not \NP-complete.

A natural question is whether {\sc Dual} lies in some lower complexity
class based on other resources than just runtime.  In the present
paper, we advance the complexity status of this problem by showing
that its complement is feasible with {\em limited nondeterminism},
i.e, by a nondeterministic polynomial-time algorithm that makes only a
poly-logarithmic number of guesses. For a survey on complexity classes
with limited nondeterminism, and for several references
see~\cite{gold-etal-96}.  
We first show by using a simple but
effective technique, which succinctly describes computation paths,
that testing non-duality is feasible in polynomial time with $O(\log^3
n)$ nondeterministic steps. We then observe that this approach can be
improved to obtain a bound of $O(\chi(n)\cdot\log n)$ nondeterministic
steps. {\em This result is surprising, because most researchers
dealing with the complexity of {\sc Dual} and {\sc Trans-Hyp} believed
so far that these problems are completely unrelated to limited
nondeterminism.}

We believe that the results presented in this paper are significant,
and we are confident that they will be prove useful in various
contexts. First, we hope that the various
polynomial/output-polynomial cases of the problems which we identify
will lead to better and more general methods in various application
areas (as we show, e.g.\ in learning and data
mining~\cite{domingo-etal-99}), and that based on the algorithm {\sc
Dualize} or some future modifications, further relevant tractable
classes will be identified. Second, we hope that our discovery on
limited nondeterminism provides a new momentum to complexity research
on {\sc Dual} and {\sc Trans-Hyp}, and will push it towards settling
these longstanding open problems.

The rest of this paper is structured as follows. The next section
provides some preliminaries and introduces notation. In
Section~\ref{sec:ordered}, we present our algorithm {\sc Dualize} for
dualizing a given monotone prime CNF. After that, we exploit this
algorithm in Section~\ref{sec:polynomial-cases} to derive a number of
polynomial instance classes of the problems {\sc Dualization} and {\sc
Dual}. In Section~\ref{sec:nondet} we then show that {\sc Dual} can be
solved with limited nondeterminism.

\section{Preliminaries and Notation}
\label{sec:prelim}

A {\em Boolean function} (in short, {\em function}) is a mapping
$f: \{0,1\}^{n} \to \{0,1\}$,
where $v \in \{0, 1\}^n$ is called a {\em Boolean vector} (in short, {\em
vector}).
As usual, we write  $g \leq f$ if $f$ and $g$ satisfy $g(v)
\leq f(v)$ for  all $v \in \{0,1 \}^{n}$, and $g<f$ if $g\leq f$ and
$g\neq f$.
A function $f$ is {\em monotone} (or {\em positive}), if $v \leq w$
(i.e., $v_i \leq w_i$ for all $i$) implies $f(v) \leq f(w)$ for all
$v,w \in \{0,1\}^n$.  Boolean variables $x_1, x_2, \ldots , x_n$ and their
complements $\bar{x}_1, \bar{x}_2, \ldots , \bar{x}_n$ are called {\em
literals}.  A {\em clause} (resp., {\em term}) is a disjunction
(resp., conjunction) of literals containing at most one of $x_i$ and
$\bar{x}_i$ for each variable.  A clause $c$ (resp., term $t$) is an
{\em implicate} (resp., {\em implicant}) of a function $f$, if $f \leq
c$ (resp., $t \leq f$); moreover, it is {\em prime}, if there is no
implicate $c' < c$ (resp., no implicant $t' > t$) of $f$, and {\em
monotone}, if it consists of positive literals only. We denote by
$PI(f)$ the set of all prime implicants of $f$. 

A {\em conjunctive normal form} (CNF) (resp., disjunctive normal form,
DNF) is a conjunction of clauses (resp., disjunction of terms); it is
{\em prime} (resp.\ {\em monotone}), if all its members are prime
(resp.\ {\em monotone}). For any CNF (resp., DNF) $\rho$, we denote by
$|\rho|$ the number of clauses (resp., terms) in it. Furthermore, for
any formula $\gvp$, we denote by $V(\gvp)$ the set of variables that
occur in $\gvp$, and by $\| \gvp \|$ its {\em length}, i.e.,
the number of literals in it.
 We occasionally view CNFs $\gvp$ also as sets of clauses, and
clauses as sets of literals, and use respective notation (e.g., $c\in
\gvp$, $\ol{x_1}\in c$ etc). 

As well-known, a function $f$ is monotone iff it has a monotone
CNF. Furthermore, all prime implicants and prime implicates of a
monotone $f$ are monotone, and it has a unique prime CNF, given by the
conjunction of all its prime implicates.  For example, the monotone
$f$ such that $f(v)=1$ iff $v\in
\{ (1100), (1110), (1101), (0111), (1111)\}$ has the unique prime CNF
$\gvp= x_2(x_1 \vee x_3)(x_1 \vee x_4)$.

Recall that the {\em dual} of a function $f$, denoted $f^{d}$, is
defined by
$f^{d}(x) = \ol{f}(\ol{x})$,
where $\ol{f}$ and $\ol{x}$ is the complement of $f$ and $x$,
respectively.  By definition, we have $(f^d)^d=f$. From De Morgan's
law, we obtain a formula for $f^{d}$ from any one of $f$ by exchanging
$\vee$ and $\wedge$ as well as the constants $0$ and $1$. For example,
if $f$ is given by $\gvp=x_1x_2 \vee \ol{x}_1(\ol{x}_3 \vee x_4)$,
then $f^d$ is represented by $\psi=(x_1 \vee x_2)(\ol{x}_1 \vee
\ol{x}_3x_4)$. For a monotone function $f$, let $\psi=\AN_{c \in C}( \OR_{x_i
\in c}x_i)$ be the prime CNF of $f^d$.  Then by De Morgan's
law, $f$ has the (unique) prime DNF $\rho=\OR_{c \in C}(\AN_{x_i \in
c}x_i)$; in the previous example, $\rho = x_1 x_2 \lor x_2
x_3x_4$. Thus, we will regard {\sc Dualization} also as the problem of
computing the prime DNF of $f$ from the prime CNF of $f$.

\section{Ordered Transversal Generation}
\label{sec:ordered}

In what follows, let $f$ be a monotone function and
\begin{eqnarray}
\gvp &=&\AN_{i=1}^mc_i \label{eq-1}
\end{eqnarray}
its prime CNF, where we assume
without loss of generality 
that all variables $x_j$ ($j=1,2, \ldots n$) appear in $\gvp$.  Let
$\gvp_i$ ($i=0,1, \ldots , n$) be the CNF obtained from $\gvp$ by
fixing variables $x_j=1$ for all $j$ with $j \geq i+1$.  By
definition, we have $\gvp_0= 1$ (truth) and $\gvp_n=\gvp$. For
example, consider $\gvp=(x_1 \vee x_2)(x_1 \vee x_3)(x_2 \vee x_3 \vee
x_4)(x_1 \vee x_4)$.  Then we have $\gvp_0=\gvp_1=1$, $\gvp_2=(x_1
\vee x_2)$, $\gvp_3=(x_1 \vee x_2)(x_1 \vee x_3)$, and $\gvp_4=\gvp$.
Similarly, for the prime DNF
\begin{eqnarray}
\psi&=&\textstyle \OR_{t \in \gG(f)}t  \label{eq-2}
\end{eqnarray}
of $f$, we denote
by $\psi_i$ the DNF obtained from $\psi$ by fixing variables $x_j=1$
for all $j$ with $j \geq i+1$.  Clearly, we have $\gvp_i \equiv
\psi_i$, i.e., $\gvp_i$ and $\psi_i$ represent the same function
denoted by $f_i$. 

\begin{proposition}
\label{prop1}
Let $\gvp$ and $\psi$ be any CNF and DNF for $f$, respectively. Then,
for all $i\geq 0$, 
\begin{itemize}
    \item[(a)] 
$\|\gvp_i\| \leq \|\gvp\|$ and  $|\gvp_i| \leq |\gvp|$, and 
\item[(b)]
$\|\psi_i\| \leq \|\psi\|$ and  $|\psi_i| \leq |\psi|$. 
\end{itemize}
\end{proposition}
Denote by $\gD^i$ ($i=1,2, \ldots ,n$) the CNF consisting of all the clauses
in $\gvp_i$ but not in $\gvp_{i-1}$.
For the above example, we have $\gD^1=1$, $\gD^2=(x_1 \vee x_2)$,
$\gD^3=(x_1
\vee x_3)$, and
$\gD^4=(x_2 \vee x_3 \vee x_4)(x_1 \vee x_4)$.
Note that $\gvp_{i} =\gvp_{i-1} \wedge \gD^{i}$; hence,
for all $i=1,2,\ldots,n$ we have
\begin{eqnarray}
\psi_{i}&\equiv& \psi_{i-1} \wedge \gD^{i}
\ \equiv \
\OR_{t \in \gG(f_{i-1})} (t \wedge \gD^{i}). \label{eq-3}
\end{eqnarray}
Let $\Delta^{i}[t]$, for $i=1,\ldots,n$ denote the CNF consisting
of all the clauses $c$ such that $c$ contains no literal in $t_{i-1}$ and
$c \vee x_{i}$ appears in $\Delta^{i}$.  For example, if
$t=x_2x_3x_4$ and $\gD^4=(x_2 \vee x_3 \vee x_4)(x_1
\vee x_4)$, then $\gD^4[t]=x_1$.
It follows from \raf{eq-3}  that for all $i=1,2,\ldots,n$ 
\begin{eqnarray}
\psi_{i}&\equiv&  \OR_{t \in \gG(f_{i-1})} \Bigl((t\wedge\gD^{i}[t]) \ \vee\
(t\wedge x_{i})\Bigr). \label{eq-4}
\end{eqnarray}
\begin{lemma}
\label{lemma-1}
For every term $t\in \gG(f_{i-1})$, let $g_{i,t}$ be the function represented by
$\gD^{i}[t]$. Then $|\gG(g_{i,t})|\leq|\psi_{i}|\leq|\psi|$.
\end{lemma}

\proof   Let $V =\{x_1,x_2, \ldots , x_n\}$ and let $s \in \gG(g_{i,t})$.
Then by \raf{eq-4}, $t \wedge s$ is an implicant of $\psi_{i}$.
Hence, some $t^s \in \gG(f_{i})$ exists such
that $t^s \geq t \wedge s$.
Note that
$V(t) \cap V(\gD^{i}[t])=\emptyset$, 
$t$ and $\gD^{i}[t]$ have no variable in common,
and hence we have $V(s) \subseteq V(t^s) \,(\subseteq V(s) \cup
V(t))$, 
since otherwise there exists a clause $c$ in $\gD^{i}[t]$
such that $V(c) \cap V(t^s) =\emptyset$, a contradiction.  
Thus $V(t^s)\cap V(\gD^{i}[t]) = V(s)$.  
For any $s' \in \gG(g_{i,t})$ such that $s \neq s'$, 
let $t^s, t^{s'} \in \gG(f_{i})$ such
that $t^s \geq t \wedge s$ and $t^{s'} \geq t \wedge s'$, respectively. 
By the above discussion, we have $t^s \neq t^{s'}$.  
This completes the proof. \qed

We now describe our algorithm {\sc Dualize} for generating $\gG(f)$.
It is inspired by a similar graph algorithm of Johnson, Yannakakis, and
Papadimitriou \cite{john-etal-88}, and can be regarded as a generalization.

\vspace*{0.5cm}

\hrule
\begin{small}
\begin{quote}
\noindent {\bf Algorithm} {\sc Dualize}

\hnbls
\tnbls

\begin{tabbing}
{\em Output: }\=\kill
{\em Input:}\> The prime CNF $\gvp$ of a monotone function $f$.
\\
{\em Output:}\> The prime DNF $\psi$ of $f$, i.e. all prime
implicants
of $f$.
\end{tabbing}

\nbls
\tnbls

\begin{description}
\addtolength{\itemsep}{-0.1cm}
\item[\hspacea Step 1:] Compute the smallest prime implicant $t_{min}$ of
$f$ and set $Q := \{\, t_{min} \,\}$;

\item[\hspacea Step 2:] {\bf while} \ $Q\not=\emptyset$ \ {\bf do}
{\bf  begin}

\hspacec Remove the smallest $t$ from $Q$ and output $t$;

\hspacec {\bf for} each $i$ with $x_i \in V(t)$ and
						$\gD^{i}[t]\not= 1$ {\bf do begin}



\hspaced Compute the prime DNF $\rho_{(t,i)}$ of the function
						represented by $\gD^{i}[t]$;

\hspaced {\bf for} each term $t'$ in $\rho_{(t,i)}$ {\bf do begin}

\hspacee {\bf if} $t_{i-1} \wedge t'$ is a prime implicant of $f_i$
						{\bf then begin}

\hspacef Compute the smallest prime implicant $t^*$ of $f$ such that
						$t^*_i=t_{i-1} \wedge t'$;

\hspacef $Q:=Q \cup \{t^*\}$

\hspacee {\bf end}$\{$if$\}$\ {\bf end}$\{$for$\}$\ {\bf end}$\{$for$\}$

\hspaceb {\bf end}$\{$while$\}$
\end{description}
\end{quote}
\end{small}
\hrule

\vspace*{0.5cm}

\noindent
Here,
we say that term $s$ is {\em smaller} than term $t$ if
$\sum_{x_j \in V(s)}2^{n-j} < \sum_{x_j \in V(t)}2^{n-j}$; i.e., as
vector, $s$ is lexicographically smaller than $t$.

\begin{theorem}
\label{th-1}
Algorithm {\sc Dualize} correctly outputs all $t\in \gG(f)$ in
increasing order.
\end{theorem}

\proof  (Sketch)
First note that the term $t^*$ inserted in $Q$ when $t$ is output is
larger than $t$. Indeed, $t'$ ($\neq 1$) and $t_{i-1}$ are disjoint
and $V(t') \subseteq \{x_1,$$\ldots,x_{i-1}\}$. Hence, every term in $Q$
is larger than all terms already output, and the
output sequence is increasing.  We show by induction that, if
$t$ is the smallest prime implicant of $f$ that was not output yet,
then $t$ is already in $Q$.  This clearly proves the result.

Clearly, the above statement is true if $t=t_{min}$. Assume now that
$t \neq t_{min}$ is the smallest among the prime implicants not output
yet.  Let $i$ be the largest index such that $t_i$ is not a prime
implicant of $f_i$.  This $i$ is well-defined, since otherwise $t =
t_{min}$ must hold, a contradiction.  Now we have (1) $i < n$ and (2)
$i+1 \not\in V(t)$, where (1) holds because $t_n\,(=t)$ is a prime
implicant of $f_n\,(=f)$ and (2) follows from the maximality of $i$.
Let $s \in \gG(f_i)$ such that $V(s) \subseteq V(t_i)$, and let $K=V(t_i)-V(s)$.
Then $K\not=\emptyset$ holds, and since $x_{i+1}\notin V(t)$, the term
$t'=\AN_{x_j \in K}x_j$ is a prime implicant of $\gD^{i+1}[s]$.  There
exists $s' \in \gG(f)$ such that $s'_i=s$ and $x_{i+1}
\in V(s')$, since $s \wedge x_{i+1} \in \gG(f_{i+1})$.
Note that $\gD^{i+1}[s]\not= 0$.  Moreover, since
$s'$ is smaller than $t$, by induction $s'$ has already been output.
Therefore, $t'=\AN_{x_j \in K}x_j$ has been considered in the inner
for-loop of the algorithm.  Since $s_i' \wedge t' \,(=t_i=t_{i+1})$ is
a prime implicant of $f_{i+1}$, the algorithm has added the smallest
prime implicant $t^*$ of $f$ such that $t^*_{i+1}=t_{i+1}$.  We
finally claim that $t^*=t$. Otherwise, let $k$ be the first index in
which $t^*$ and $t$ differ.  Then $k > i+1$, $x_k \in V(t)$ and $x_k
\not\in V(t^*)$.  However, this implies $t_k \notin \gG(f_k)$, contradicting
the maximality of $i$.
\qed

\begin{remark}
{\rm (1) The decomposition rule \raf{eq-4}
was already used  in \cite{lawl-etal-80}. 

\noindent (2) In step~1, we could generate any prime implicant $t$ of
$f$, and choose then a lexicographic term ordering inherited from a
dynamically generated variable ordering. In step~2, it is sufficient
that any monotone DNF $\tau_{(t,i)}$ of the function represented by
$\gD^{i}[t]$ is computed,
rather than its prime DNF $\rho_{(t,i)}$. This might make the
algorithm faster. 
}
\end{remark}


Let us consider the time complexity of algorithm {\sc Dualize}. We store $Q$
as a
binary tree, where each leaf represents a term $t$ and
the left (resp., right) son of a node at depth $j-1\geq 0$, where the
root has depth 0, encodes $x_j \in V(t)$ (resp., $x_j \not\in V(t)$).
In Step 1, we can compute $t_{min}$ in
$O(\|\gvp\|)$ time and initialize $Q$ in $O(n)$ time.

As for Step 2, let $T_{(t,i)}$  be the time required to compute the prime
DNF
$\rho_{(t,i)}$ from $\gD^{i}[t]$. By analyzing its substeps, we can
        see that each iteration of Step 2 requires
$\sum_{x_i \in V(t)}(T_{(t,i)}+|\rho_{(t,i)}|\cdot O(\|\gvp\|))$
time.

Indeed, we can update $Q$ (i.e., remove the smallest term and add
$t^*$) in $O(n)$ time.  For each $t$ and $i$, we can construct
$\gD^{i}[t]$ in $O(\| \gvp\|)$ time.  Moreover, we can check whether
$t_{i-1} \wedge t'$ is a prime implicant of $f_i$ and if so, we can
compute the smallest prime implicant $t^*$ of $f$ such that
$t^*_i=t_{i-1} \wedge t'$ in $O(\|\gvp\|)$ time; note that $t^*$ is
the smallest prime implicant of the function
obtained from $f$ by fixing $x_j=1$ if $x_j \in V(t_i \wedge t')$ and
$0$ if $x_j \not\in V(t_i \wedge t')$ for $j \leq i$.

Hence, we have the following result.
\begin{theorem}
\label{th-2}
The output delay  of Algorithm {\sc Dualize} is bounded by
\begin{equation}
\label{eq-6}
\max_{t \in \gG(f)}\Bigl(\sum_{x_i \in
V(t)}(T_{(t,i)}+|\rho_{(t,i)}|\cdot O(\|\gvp\|))\Bigr)
\end{equation}
time, and {\sc Dualize} needs in total time
\begin{equation}
\label{eq-7}
\sum_{t \in \gG(f)}\sum_{x_i \in
V(t)}(T_{(t,i)}+|\rho_{(t,i)}|\cdot O(\|\gvp\|)).
\end{equation}
\end{theorem}

If the $T_{(t,i)}$ are bounded by a polynomial in the input length,
 then {\sc Dualize} becomes a polynomial delay
algorithm, since $|\rho_{(t,i)}| \leq T_{(t,i)}$ holds for all $t \in\gG(f)$
and $x_i \in V(t)$.
On the other hand, if they are bounded by a polynomial in the combined
input and output length,
 then {\sc Dualize} is a polynomial total time algorithm,
where $|\rho_{(t,i)}| \leq |\psi|$ holds from Lemma~\ref{lemma-1}.
Using results from \cite{bioc-ibar-95}, we can construct
from {\sc Dualize} an incremental polynomial time algorithm for {\sc
Dualization}, which however might not output $\gG(f)$ in increasing
order.
Summarizing, we have the following corollary.
\begin{corollary}
\label{cor-0} Let $T=\max \{\, T_{(t,i)} \mid t \in \gG(f), x_i \in V(t) \,\}$. Then,
if $T$ is bounded by a
\begin{itemize}
\item[(i)]
polynomial in $n$ and $\|\gvp\|$, then {\sc
Dualize} is an $O(n
\|\gvp\| T)$ polynomial delay algorithm; 
\item[(ii)]
polynomial in $n$, $\|\gvp\|$, and $\|\psi\|$,
then {\sc Dualize} is an $O(n\cdot|\psi|\cdot(T+|\psi|\cdot\|\gvp\| ))$ polynomial
total-time algorithm; moreover, {\sc Dualization}
is solvable in incremental polynomial time.
\end{itemize}
\end{corollary}

In the next section, we identify sufficient conditions for the boundedness
of $T$ and fruitfully apply them to solve open problems and improve
previous results.

\section{Polynomial Classes}
\label{sec:polynomial-cases}

\subsection{Degenerate CNFs}

We first consider the case of small $\Delta^i[t]$.  Generalizing a
notion for graphs (i.e., monotone $2$-CNFs) \cite{toft-95}, we call a
monotone CNF $\gvp$ {\em $k$-degenerate}, if there exists a variable
ordering $x_1,\ldots,x_n$ in which $|\Delta^i| \leq k$ for all $i=1,2,
\ldots, n$. We call a variable ordering $x_1,\ldots,x_n$ {\em smallest
last} as in \cite{toft-95}, if $x_i$ is chosen in the order $i=n,n-1,
\ldots , 1$ such that $|\Delta^i|$ is smallest for all variables
that were not chosen. Clearly, a smallest last ordering gives the
least $k$ such that $\gvp$ is $k$-degenerate.  Therefore, we can check
for every integer $k\geq 1$ whether $\gvp$ is $k$-degenerate in
$O(\|\gvp\| )$ time.  If this holds, then we have $|\rho_{(t,i)}| \leq n^k$
and $T_{(t,i)}=O(kn^{k+1})$ for every $t \in \gG(f)$ and $i \in V(t)$
(for $T_{(t,i)}$, apply the distributive law to $\Delta^i[t]$ and
remove terms $t$ where some $x_j\in V(t)$ has no $c\in
\Delta^i[t]$ such that $V(t)\cap V(c)=\{x_j\}$).  Thus Theorem~\ref{th-2}
implies the following.
\begin{theorem}
\label{th-3}
For $k$-degenerate CNFs $\gvp$, 
{\sc Dualization} is solvable with $O(\|\gvp \|\cdot n^{k+1})$ polynomial
delay if $k\geq 1$ is constant.
\end{theorem}

Applying the result of \cite{maki-00} that log-clause CNF is dualizable
in incremental polynomial time,  we obtain a polynomiality result also
for non-constant degeneracy:
\begin{theorem}
\label{th-4}
For  $O(\log \|\gvp\| )$-degenerate CNFs $\gvp$, problem {\sc
Dualization} is solvable in polynomial total time.
\end{theorem}

In the following, we discuss several natural subclasses of degenerate CNFs.

\subsubsection{Read-bounded CNFs}

A monotone CNF $\gvp$ is called {\em read-$k$}, if each
variable appears in $\gvp$ at most $k$ times.
Clearly, read-$k$ CNFs are $k$-degenerate, and in fact $\gvp$ is
read-$k$ iff it is $k$-degenerate under every variable ordering.
By applying Theorems~\ref{th-3} and \ref{th-4}, we obtain the following
result.
\begin{corollary}
\label{cor-3}
For read-$k$ CNFs  $\gvp$, problem {\sc Dualization} is solvable 
\begin{itemize}
\item[(i)]
with
$O(\|\gvp\|\cdot n^{k+1})$ polynomial delay, if $k$ is constant;
\item[(ii)]
in polynomial total time, if $k=O(\log(\|\gvp\|))$.
\end{itemize}
\end{corollary}

Note that Corollary~\ref{cor-3} (i) trivially implies that {\sc
Dualization} is solvable in $O(|\psi|\cdot n^{k+2})$ time for constant
$k$, since $\|\gvp\|\leq kn$.  This improves upon the
previous best known algorithm \cite{domingo-etal-99}, which is only
$O(|\psi|\cdot n^{k+3})$ time, not polynomial delay, and outputs $\gG(f)$
in no specific order.  Corollary \ref{cor-3} (ii) is a non-trivial
generalization of the result in \cite{domingo-etal-99}, which was
posed as an open problem \cite{domingo-97}.

\subsubsection{Acyclic CNFs}

Like in graphs, acyclicity is appealing in hypergraphs resp.\ monotone
                    CNFs from a theoretical as well as a practical
                    point of view.
However, there are many notions of acyclicity for hypergraphs (cf.\
\cite{fagi-83}), since different generalizations
from graphs are possible. We refer to $\ga$-, $\gb$-,$\gc$-, and {\em Berge}-acyclicity
as stated
in \cite{fagi-83}, for which the following proper inclusion hierarchy is
known:
\begin{center}
Berge-acyclic $\,\subseteq\,$ $\gc$-acyclic $\,\subseteq\,$ $\gb$-acyclic
$\,\subseteq\,$ $\ga$-acyclic.
\end{center}
The notion of $\ga$-acyclicity came up in  relational
database theory. A monotone CNF $\gvp$ is {\em $\ga$-acyclic} iff $\gvp=1$
or reducible by
the GYO-reduction \cite{grah-79,yu-ozsoyoglu-79}, i.e.,
repeated application of one of the two rules:

\medskip
(1)~~ If variable $x_i$ occurs in only one clause $c$, remove $x_i$ from
$c$.

(2)~~ If distinct clauses $c$ and $c'$ satisfy $V(c) \subseteq
    V(c')$, remove $c$ from $\gvp$.

\medskip
\noindent
to $0$ (i.e., the empty clause).  Note that $\ga$-acyclicity of a
monotone CNF $\gvp$ can be checked, and a suitable GYO-reduction
output, in $O(\|\gvp\| )$ time \cite{tarjan-yannakakis-84}.
A monotone CNF $\gvp$ is {\em $\gb$-acyclic} iff every CNF consisting
of clauses in $\gvp$ is $\ga$-acyclic. As shown in
\cite{eite-gott-95}, the prime implicants of a monotone $f$
represented by a $\gb$-acyclic CNF $\gvp$ can be enumerated (and thus
{\sc Dualization} solved) in $p(\|\gvp\|)\cdot|\psi|$ time, where $p$ is a
polynomial in $\|\gvp\|$. However, the time complexity of {\sc
Dualization} for the more general $\ga$-acyclic prime CNFs was left as
an open problem.  We now show that it is solvable with polynomial
delay, by showing that $\ga$-acyclic CNFs are $1$-degenerate.

Let $\gvp \neq 1$ be a prime CNF. Let $a = a_1, a_2, \ldots , a_q$ be
a GYO-reduction for $\gvp$, where
$a_\ell=x_i$ if the $\ell$-th operation removes $x_i$ from $c$, and
$a_\ell=c$ if it removes $c$ from $\gvp$.
Consider the unique variable ordering $b_1, b_2,
\ldots , b_n$ such $b_i$ occurs after $b_j$ in $a$, for all $i<j$.
For example, let $\gvp=c_1c_2c_3c_4$,
where
$c_1=(x_1 \vee x_2 \vee x_3)$, $c_2=(x_1 \vee x_3 \vee x_5)$, $c_3=(x_1
\vee x_5 \vee x_6)$ and $c_4=(x_3 \vee x_4 \vee x_5)$.
Then $\gvp$ is $\ga$-acyclic, since it has the GYO-reduction
$$
a_1=x_2, \ a_2=c_1, \ a_3=x_4, \ a_4=x_6,\  a_5=c_4,\  a_6=c_3,
\  a_7=x_1, \ a_8=x_3, \  a_9=x_5.
$$
{From} this sequence, we obtain the variable ordering
$$b_1=x_5, \ b_2=x_3, \ b_3=x_1,  \ b_4=x_6, \ b_5=x_4, \ b_6=x_2.$$
As easily checked, this ordering shows that $\gvp$ is $1$-degenerate.
Under this ordering,
we have $\Delta^1=\Delta^2=1$, $\Delta^3=(x_1 \vee x_3 \vee x_5)$,
$\Delta^4=(x_1 \vee x_5 \vee x_6)$,  $\Delta^5=(x_3 \vee x_4 \vee x_5)$,
and  $\Delta^6=(x_1 \vee x_2 \vee x_3)$. This is not accidental.
\begin{lemma}
\label{lemma-2}
Every $\ga$-acyclic prime CNF is $1$-degenerate.
\end{lemma}

Note that the converse is not true, i.e., there exists a $1$-degenerate CNF that is not $\ga$-acyclic.
For example, $\gvp=(x_1 \vee x_2 \vee x_3)(x_1 \vee x_2 \vee x_4)(x_2\vee
x_3 \vee x_4 \vee x_5)$ is such a CNF.
Lemma~\ref{lemma-2} and Theorem~\ref{th-3} imply the following
result.
\begin{corollary}
\label{cor-2}
For $\ga$-acyclic CNFs $\gvp$, problem {\sc Dualization} is solvable with
$O(\|\gvp\|\cdot n^{2})$ delay.
\end{corollary}
Observe that for a prime $\alpha$-acyclic $\gvp$, we have $|\gvp|\leq
n$. Thus, if we slightly modify algorithm {\sc Dualize} to check
$\Delta^i=1$ in advance (which can be done in linear time in a
preprocessing phase) such that such $\Delta^i$ need not be considered
in step 2, then the resulting algorithm has $O(n\cdot|\gvp|\cdot\|\gvp\|)$
delay.  Observe that the algorithm in \cite{eite-gott-95} solves,
minorly adapted for enumerative output, {\sc Dualization} for
$\beta$-acyclic CNFs with $O(n\cdot|\gvp|\cdot\|\gvp\|)$ delay. Thus, the above
modification of {\sc Dualize} is of the same order.


\subsubsection{CNFs with bounded treewidth}

A {\em tree decomposition (of type I)} of a monotone CNF $\gvp$ is a
tree $T=(W,E)$ where each node $w\in W$ is labeled with a
set $X(w)\subseteq V(\gvp)$ under the following conditions:
\begin{enumerate}

\item $\bigcup_{w \in W}X(w) =V(\gvp)$;

\item for every clause $c$ in $\gvp$, there exists some $w\in W$ such that
$V(c) \subseteq X(w)$; and

\item for any variable $x_i \in V$, the set of nodes $\{w \in W \mid x_i \in X(w)\}$ induces
a (connected) subtree of $T$.
\end{enumerate}
The {\em width} of $T$ is $\max_{w \in W}|X(w)|-1$, and the {\em
treewidth} of $\gvp$, denoted by $\Tw_1(\gvp)$, is the minimum width
over all its tree decompositions.

Note that the usual definition of treewidth for a graph \cite{robe-seym-86} results in the
case where $\gvp$ is a 2-CNF. Similarly to  acyclicity, there are several
notions of treewidth for hypergraphs resp.\ monotone CNFs. For example, tree
decomposition of type II of CNF $\gvp=\AN_{c \in C}c$ is
defined as type-I tree decomposition of its incident $2$-CNF (i.e.,
graph) $G(\gvp)$
\cite{chekuri-rajaraman-95,gottlob-etal-98}.
That is, for each clause $c \in \gvp$, we introduce a new variable $y_c$ and
construct $G(\gvp)=\AN_{x_i \in c \in \gvp} (x_i \vee y_c)$
(here, $x_i \in c$ denotes that $x_i$ appears in $c$).
Let $\Tw_2(\gvp)$ denote the type-II treewidth of $\gvp$.

\begin{proposition}
\label{prop-2}
For every monotone CNF $\gvp$, it holds that $\Tw_2(\gvp)\leq
\Tw_1(\gvp)+2^{\Tw_1(\gvp)+1}$.
\end{proposition}

\proof
Let $T=(W,E)$, $X:W \to 2^V$ be any tree decomposition of $\gvp$
having width $\Tw_1(\gvp)$.
Introduce for all $c \in \gvp$ new variables $y_c$, and
add $y_c$ to every $X(w)$ such that $V(c) \subseteq X(w)$.  Clearly, the
result is a type-I tree
decomposition of $G(\gvp)$,
and thus a type-II tree decomposition of $\gvp$.
Since at most $2^{|X(w)|}$ many $y_c$ are added to $X(w)$ and
$|X(w)|-1 \leq \Tw_1(\gvp)$ for every $w \in W$, the result follows.
\qed

This means that if $\Tw_1(\gvp)$ is bounded by some constant, then so
is $\Tw_2(\gvp)$. Moreover, $\Tw_1(\gvp)=k$ implies that $\gvp$ is a
$k$-CNF; we discuss $k$-CNFs in Section~\ref{sec:recursive} and only
consider $\Tw_2(\gvp)$ here. The following proposition states some
relationships between type-II
treewidth and other restrictions of CNFs from above.

\begin{proposition}
The following properties hold for type-II treewidth.
\begin{enumerate}
\item[(i)]
There is a family of monotone prime CNFs $\gvp$ such that
$\Tw_2(\gvp)$ is bounded by a constant, but $\gvp$ is not $k$-CNF for any
constant $k$.

\item[(ii)] There is a family of monotone prime CNFs $\gvp$
such that 
$\Tw_2(\gvp)$ is bounded by a constant, but $\gvp$ does not have bounded
read.

\item[(iii)] There is a family of $\ga$-acyclic
prime CNFs $\gvp$ such that $\Tw_2(\gvp)$ is not bounded by any
constant. (This is a contrast to the graph case that a graph is
acyclic if and only if its treewidth is $1$.) 
\end{enumerate}
\end{proposition}

\proof
(i): For example, $\gvp=(\OR_{x_i \in V}x_i)$ has $\Tw_2(\gvp)=1$, since
it has a tree decomposition $T=(W,E)$ with $X: W \to 2^V$ defined by
$W=\{1,2, \ldots ,n \}$, $E=\{(w,w+1), w=1,2, \ldots , n-1\}$, and
$X(w)=\{x_w,y_c\}$, $w \in W$, where $c=(\OR_{x_i \in V}x_i)$.
However, it is not an $(n-1)$-CNF (but an $n$-CNF).  On the other hand,
by Lemma \ref{lemma-3}, we can see that there is a family of monotone
prime CNFs $\gvp$ such that $\Tw_2(\gvp)$ is not bounded by any
	   constant,
but $\gvp$ is $k$-CNF for some constant $k$.

(ii): For example, let $\gvp$ be a CNF containing $n-1$ clauses
$c_i=(x_1 \vee x_i)$, $i=2,3, \ldots ,n$.  Then $\gvp$ has
$\Tw_2(\gvp)=1$, since it has a tree decomposition $T=(W,E)$ with $X:
W \to 2^V$ defined by $W=\{(c_i,x_1), (c_i,x_i), i=2,3, \ldots ,n \}$,
$E=\{((c_i,x_1),(c_{i+1},x_1)), i=2,3, \ldots , n-1\} \cup
\{((c_i,x_1),(c_{i},x_i)), i=2,3, \ldots , n$, and $X((c_i,x_k))=
\{y_{c_i},x_k\}$, $(c_i,x_k) \in W$.  However, it is not read-($n-2)$
(but read-$(n-1)$).

(iii): For example, let $\gvp$
be a CNF on $V=\{x_1,x_2, \ldots, x_{2n}\}$ containing $n$ clauses
$c_i=(x_i \vee \OR_{j\geq n+1}x_j)$, for $i=1,\ldots,n$.  Then $\gvp$
is $\ga$-acyclic.  We claim that $\Tw_2(\gvp) \geq n-1$.  Let us
assume that there exists a tree $T=(W,E)$ with $X: E \to 2^V$ that
shows  $\Tw_2(\gvp) \leq n-2$, where $T$ is regarded as a rooted
tree.  Let $T_i=(W_i,E_i)$ be the subtree of $T$ induced by $W_i=\{w
\in W \mid y_{c_i} \in X(w)\}$, and let $r_i$ be its root.  Consider
the case in which $W_i$ and $W_{j}$ are disjoint for some $i$ and $j$.
Suppose that $r_{j}$ is an ancestor of $r_i$.  Since $|X(r_i)| \leq
\Tw_2(\gvp)+1 \leq n-1$, there exists a node $x_{n+k}\in V$ such
that $1 \leq k \leq n$ and $x_{n+k} \not\in X(r_i)$.  However, since
the incident graph of $\gvp$ contains two edges $(x_{n+k},y_{c_i})$
and $(x_{n+k},y_{c_j})$, we have $x_{n+k} \in \bigcup_{w \in W_i
-\{r_i\}}X(w)$ and $x_{n+k} \in \bigcup_{w \in W_{j}}X(w)$. This is
a contradiction to the condition that $\{w \in W \mid x_{n+k} \in
X(w)\}$ is connected.  Similarly, we can prove our claim when $T_i$
and $T_{j}$ are disjoint, but $r_{j}$ is not an ancestor of $r_i$.

We thus consider the case in which $W_i \cap W_{j} \not= \emptyset$
holds for any $i$ and $j$. Since $T_i$'s are trees, the family of
$W_i$, $i=1,2, \ldots, n$, satisfies the well-known Helly property,
i.e., there exists a node $w$ in $\bigcap_{i=1}^n W_i$.  $X(w)$
must contain all $y_{c_i}$'s.  This implies $|X(w)| \geq n$, a
contradiction.
\qed

As we show now, bounded-treewidth implies bounded degeneracy.
\begin{lemma}
\label{lemma-3}
Let $\gvp$ be any monotone CNF with $\Tw_2(\gvp) = k$.
Then $\gvp$ is $2^k$-degenerate.
\end{lemma}

\proof 
Let $T=(W,E)$ with $X: W \to 2^V$ show  $\Tw_2(\gvp) =k$.
{From this}, we reversely construct a variable ordering $a=a_1,\ldots,a_n$
on $V=V(\gvp)$ such
that $|\gD^i| \leq 2^k$ for all $i$.

Set $i:=n$. Choose any leaf $w^*$ of $T$, and let $p(w^*)$ be a node
in $W$ adjacent to $w^*$. If $X(w^*)\setminus X(p(w^*)) \subseteq
\{y_c \mid c \in \gvp\}$, then remove $w^*$ from $T$. On the other
hand, if $(X(w^*)\setminus X(p(w^*))) \cap V = \{x_{j_1},\ldots,
x_{j_\ell}\}$ where $\ell\geq 1$ (in this case, only $X(w^*)$ contains
$x_{j_1}$, \ldots, $x_{j_\ell}$), then define $a_{i+1-h}=x_{j_h}$ for
$h=1,\ldots ,\ell$ and update $i:=n-\ell$,
$X(w^*):=X(w^*)\setminus\{x_{j_1},\ldots , x_{j_\ell}\}$, and
$X(w):=X(w)\setminus\{y_c\mid c \in \gvp, V(c)\cap
\{x_{j_1},\ldots,x_{j_\ell}\}\neq \emptyset\,\}$ for every $w \in W$.
Let $a$ be completed by repeating this process. 

We claim that $a$ shows that $|\gD^i| \leq 2^k$ for all
$i=1,\ldots,n$. To see this, let $w^*$ be chosen during this process, and assume that $a_i \in
X(w^*)\setminus X(p(w^*))$.  Then, by induction on the (reverse) construction of $a$,
we obtain that for each clause $c\in \gD^i$ we must
have either (a) $y_c \in X(w^*)$ or (b) $V(c) \subseteq X(w^*)$. 
The latter case may arise if in previous steps of the process some
descendant $d(w^*)$ of $w^*$ was removed which contains $y_c$ such
that $y_c$ does not occur in $w^*$; however, in this case $V(c)\subseteq
X(w)$ must be true on every node on the path from $d(w^*)$ to
$w^*$. 
 
 Now let $q = |X(w^*)\setminus V|$.
Since $|X(w^*)\setminus\{a_i\}|\leq k$, we
have 
$$|\gD^i| \;\leq\; q + 2^{k-q} \;\leq\; 2^k.$$
This proves the claim.
\qed

\begin{corollary}
\label{cor-4}
For CNFs $\gvp$ with $\Tw_2(\gvp) \leq k$, {\sc Dualization} is
solvable (i) with $O(\|\gvp\|\cdot n^{2^k+1})$ polynomial delay, if $k$ is
constant; and (ii) in polynomial total time, if $k=O(\log \log \|\gvp\|)$.
\end{corollary}

\subsection{Recursive application of algorithm {\sc Dualize}}
\label{sec:recursive}

Algorithm {\sc Dualize} computes in step~2 the prime DNF
$\rho_{(t,i)}$ of the function represented by $\gD^{i}[t]$.  Since
$\gD[t]$ is the prime CNF of some monotone
function, we can recursively apply
{\sc Dualize} to $\gD^{i}[t]$ for computing $\rho_{(t,i)}$. Let us
call this variant {\sc R-Dualize}. Then we have the following result.

\begin{theorem}
\label{th-5}
If its recursion depth is $d$, {\sc R-Dualize} solves {\sc Dualization} in
$O(n^{d-1}\cdot|\psi|^{d-1}\cdot\|\gvp\|)$ time.
\end{theorem}

\proof
If $d=1$, then $\Delta^i[t_{min}]=1$ holds for $t_{min}$ and every
$i\geq 1$.  This means that $\gG(f)\!=\!\{t_{min}\}$ and $\gvp$ is a
$1$-CNF (i.e., each clause in $\gvp$ contains exactly one variable).
Thus in this case, {\sc R-Dualize} needs $O(n)$ time.  Recall that
algorithm {\sc Dualize} needs,  by \raf{eq-7}, time $\sum_{t \in
\gG(f)}\sum_{x_i \in
V(t)}(T_{(t,i)}\!+\!|\rho_{(t,i)}|\cdot O(\|\gvp\|))$.  If $d=2$, then
$T_{(t,i)}=O(n)$ and $|\rho_{(t,i)}| \leq 1$.  Therefore, {\sc R-Dualize}
needs time $O(n\cdot |\psi|\cdot\|\gvp\|)$.  For $d\geq 3$, Corollary~\ref{cor-0}.(ii) implies that {\sc R-Dualize} needs
$O(n^{d-1}\cdot|\psi|^{d-1}\cdot\|\gvp\|)$ time.
\qed

Recall that a CNF $\gvp$ is called {\em $k$-CNF} if each clause in
$\gvp$ has at most $k$ literals. Clearly, if we apply algorithm
{\sc R-Dualize} to a monotone $k$-CNF $\gvp$, the recursion depth of
{\sc R-Dualize} is at most $k$. Thus we obtain the following
result; it re-establishes, with different means, the main positive result
of \cite{BGH93,eite-gott-95}.

\begin{corollary}
\label{cor-5}
{\sc R-Dualize} solves {\sc Dualization} in
$O(n^{k-1}\cdot|\psi|^{k-1}\cdot\|\gvp\|)$ time, i.e., in polynomial total time
for monotone $k$-CNFs $\gvp$ where $k$ is constant.

\end{corollary}

\section{Limited Nondeterminism}
\label{sec:nondet}

In the previous section, we have discussed polynomial cases of
monotone dualization. In this section, we now turn to the issue of the
precise complexity of this problem. For this purpose, we consider the
decision problem {\sc Dual}, i.e., decide whether given monotone prime
CNFs $\gvp$ and $\psi$ represent dual Boolean functions, instead of
the search problem {\sc Dualization}.

It appears that problem {\sc Dual} can be solved with limited
nondeterminism, i.e., with poly-log many guessed bits by a
polynomial-time non-deterministic Turing machine. This result might
bring new insight towards settling the complexity of the problem.

We adopt Kintala and Fischer's terminology \cite{kint-fisc-84}
and write $\nondet{g(n)}$ for the class of sets accepted by a
nondeterministic Turing machine in  polynomial time making at most
$g(n)$ nondeterministic steps on every input of length $n$.
For every integer $k\geq 1$, define $\betapol{k} =
\bigcup_c\nondet{(c\log^k n)}$. The $\beta$P {\em Hierarchy} consists of the
classes
$$
\Pol=\betapol{1}\subseteq \betapol{2}\subseteq\cdots\subseteq
\bigcup_k\betapol{k}=\beta\Pol$$
and lies between \Pol\ and \NP.
The $\betapol{k}$ classes appear to be rather robust; they are closed under
polynomial time and logspace many-one reductions and have complete
problems (cf.~\cite{gold-etal-96}). The complement class of $\betapol{k}$ is
denoted by $\cobetapol{k}$.

We start in Section~\ref{subsec:a} by recalling algorithm A
of~\cite{fred-khac-96}, reformulated for CNFs and by analyzing A's
behavior. The proof that A can be converted to an algorithm that uses
$\log^3 n$ nondeterministic bit guesses, and that {\sc Dual} is thus
in $\cobetapol{3}$, is rather easy and should give the reader an
intuition of how our new method of analysis works. In
Section~\ref{subsec:b}, we use basically the same technique for
analyzing the more involved algorithm~B of ~\cite{fred-khac-96}. Using
a modification of this algorithm, we show that {\sc Dual} is in
$\cobetapol{2}$. We also prove the stronger result that the complement
of {\sc Dual} can be solved in polynomial time with only
$O(\chi(n)\cdot\log(n))$ nondeterministic steps (=bit
guesses). Finally, Section~\ref{sec:beigel-fu} shows that membership
in $\cobetapol{2}$ can alternatively be obtained by combining the
results of \cite{fred-khac-96} with a theorem of Beigel and
Fu~\cite{beig-fu-99}.

\subsection{Analysis of Algorithm A of Fredman and Khachiyan}
\label{subsec:a}

The first algorithm in \cite{fred-khac-96}  for recognizing dual
monotone pairs is as follows. 

\vspace{0.3cm}
\stepcounter{footnote}\footnotetext{In~\cite{fred-khac-96}, duality is tested for DNFs 
while our problem {\sc Dual} speaks about CNFs; this is insignificant,
since DNFs are trivially translated to CNFs for this task and vice versa (cf.\ Section~\ref{sec:prelim}).
}
\addtocounter{footnote}{-1}
\hrule
\begin{tabbing}
X\=\kill\>{\bf Algorithm} A~~(reformulated for CNFs%
\footnotemark). \\[1ex]
X \= {\em Output: }\=\kill
\>\> {\em Input:} \' Monotone CNFs $\gvp$, $\psi$ representing monotone $f$, $g\,$\ s.t.\ $V(c)\!\cap\!V(c')\!\neq\!\emptyset$, for all
$c\!\in\!\gvp$, $c'\!\in\!\psi$.\\
\>\> {\em Output:} \' {\tt yes} if $f=g^d$, otherwise a vector $w$ of form $w=(w_1,\ldots,w_m)$ such that $f(w)\neq g^d(w)$.
\\[1ex]
{\bf Step 1:}\\[0.5ex]
XX\=\kill \> Delete all redundant (i.e., non-minimal) clauses from $\gvp$ and $\psi$. 
\\[0.5ex]
{\bf Step 2:}\\[0.5ex]
\> Check that \= (1)~ $V(\phi)=V(\psi)$,~\ (2)~ $\ \max_{c\in\gvp}|c|
\leq|\psi|$,~ (3)~ $\ \max_{c'\in\psi}|c'| \leq|\gvp|$, \ and\\
\>\> (4)~ $\ \Sigma_{c\in \gvp}\, 2^{-|c|}+\Sigma_{c'\in \psi}\, 2^{-|c'|}$ $\geq 1$. \\[0.5ex]
\>  If any of
conditions (1)-(4) fails, $f\neq g^d$  and a witness $w$ is found in polynomial time (cf.~\cite{fred-khac-96}). \\[0.5ex]
{\bf Step 3:}\\[0.5ex]
\> If $|\gvp|\cdot|\psi|\leq 1$, test duality in $O(1)$ time.
\\[0.5ex]
{\bf Step 4:} \\[0.5ex]
\> If $|\gvp|\cdot|\psi|\geq 2$, find some $x_i$ occurring in  $\gvp$ or $\psi$ (w.l.o.g.\ in $\gvp$) with frequency $\geq 1/\log(|\gvp|+|\psi|)$.
\\
\> Let~ \=
\\
\>\> $\gvp_0$ = \= $\{c-\{x_i\} \mid x_i\in c,\, c \in \gvp\},$ \quad \= $\gvp_1$ = $ \{c \mid x_i\notin c,\, c\in \gvp\},$\\
\>\> $\psi_0$ = \> $ \{ c'-\{x_i\} \mid x_i\in c',\, c'\in \psi\}$, \>
$\psi_1$ =  $ \{c' \mid x_i\notin c',\, c'\in \psi\}$.
\\[0.5ex]
\> Call algorithm A on the
two pairs of forms:
\\[0.5ex]
\centerline{(A.1)~~$(\gvp_1,\, \psi_0\land \psi_1)$ \quad and \quad (A.2)~~$(\psi_1,\,
\gvp_0\land \gvp_1)$}
\\[0.5ex]
\> If both calls return {\tt yes}, then return {\tt yes} (as
   $f=g^d$), otherwise we obtain $w$ such that \\
\>  $f(w)\neq g^d(w)$ in polynomial time (cf.~\cite{fred-khac-96}).
\end{tabbing}
\hrule
\vspace{0.3cm}

We observe that, as noted in \cite{fred-khac-96}, the binary length of
any standard encoding of the input $\gvp,\psi$ to algorithm~A is
polynomially related to $|\gvp|+|\psi|$, if step~3 is reached. Thus,
for our purpose, we consider $|\gvp|+|\psi|$ to be the input size.

Let $\gvp^*$, $\psi^*$ be the original input for A. For any pair
$(\gvp,\psi)$ of CNFs, define its {\em volume} by $v=|\gvp|\cdot|\psi|$,
and let $\epsilon=1/\log n$, where $n = |\gvp^*|+|\psi^*|$.  As shown
in~\cite{fred-khac-96}, step~4 of algorithm A divides the current
(sub)problem of volume $v=|\gvp|\cdot|\psi|$ by self-reduction into
subproblems (A.1) and (A.2) of respective volumes (assuming that $x_i$
frequently occurs in $\gvp$):
\begin{eqnarray}
|\gvp_1|\cdot|\psi_0\land \psi_1|&\leq&(1-\epsilon)\cdot v\label{eq-left}\\
|\gvp_0\land \gvp_1|\cdot|\psi_1| \;\leq\; |\gvp|\cdot(|\psi|-1)&\leq &v-1 \label{eq-right}
\end{eqnarray}
Let $T=T(\gvp,\psi)$ be the recursion tree generated by A on input
$(\gvp,\psi)$. In $T$, each node $u$ is labeled with the respective
monotone pair, denoted  by $I(u)$; thus, if $r$ is the root of $T$, then
$I(r) = (\gvp,\psi)$.  The {\em  volume} $v(u)$ of node $u$ is defined 
as the volume of its label $I(u)$. 

Any node $u$ is a leaf of $T$, if algorithm~A stops on input $I(u) =
(\gvp,\psi)$ during steps 1-3; otherwise, $u$ has a left child $u_l$
and a right child $u_r$ corresponding to (A.1) and (A.2), i.e.,
labeled $(\gvp_1,\psi_0\land\psi_1)$ and $(\psi_1,\gvp_0\land\gvp_1)$
respectively. That is, $u_l$ is the ``high frequency move'' by the
splitting variable.

We observe that every node $u$ in $T$ is determined by a {\em unique path}
from the root to $u$ in $T$ and thus by a unique sequence
$seq(u)$ of right and left
moves starting from the root of $T$ and ending at $u$. The
following key lemma bounds the number of moves of each type for
certain inputs.

\begin{lemma}
\label{lem:e}
Suppose $|\gvp^*|+|\psi^*|\leq |\gvp^*|\cdot|\psi^*|$. Then for any
node $a$ in $T$, $seq(a)$ contains at most $v^*$ right moves and at most $\log^2 v^*$
left moves, where $v^*=|\gvp^*|\cdot|\psi^*|$.
\end{lemma}

\proof
By $(\ref{eq-left})$ and $(\ref{eq-right})$, each move decreases
the volume of a node label.  Thus, the length of $seq(u)$, and in
particular the number of right moves, is bounded by $v^*$. To obtain the
better bound for the left moves, we will use the following well-known
inequality:
\begin{eqnarray}
(1-1/y)^y & \leq & 1/e, \qquad \mbox{for $y\geq 1$}. \label{eq-limes}
\end{eqnarray}
In fact, the sequence $(1-1/y_i)^{y_i}$, for any $1\leq y_1<y_2<\ldots$
monotonically converges to $1/e$ from below.
By (\ref{eq-left}), the volume $v(u)$ of
any node $u$ such that $seq(u)$
contains $\log^2 v^*$ left moves is
bounded as follows:
$$
v(u) \leq v^*\cdot(1-\epsilon)^{\log^2 v^*} = v^*\cdot(1-1/\log n)^{\log^2 v^*}.
$$ 
Since $n=|\gvp^*|+|\psi^*|\leq |\gvp^*|\cdot|\psi^*|=v^*$, and because
of (\ref{eq-limes}) it follows that:
\begin{eqnarray*}
v(u)  &\leq& v^*\cdot \big( (1-1/\log v^*)^{\log v^*}\big)^{\log v^*} \\
& \leq & v^*\cdot (1/e)^{\log v^*}
\; = \; v^*/(e^{\log v^*})
\; < \; v^*/(2^{\log v^*}) \;=\;1.
\end{eqnarray*}
Thus, $u$ must be a leaf in $T$. Hence for every $u$ in $T$, $seq(u)$
contains at most
$\log^2 v^*$ left moves.
\qed

\begin{theorem}
\label{theo:beta3}
Problem {\sc Dual} is in $\cobetapol{3}$.
\end{theorem}

\proof Instances such that either $c\cap c'= \emptyset$ for some $c\in
\gvp^*$ and $c'\in \psi^*$, the sequence $seq(u)$ is empty, or
$|\gvp^*|+|\psi^*|>|\gvp^*|\cdot|\psi^*|$ are easily recognized and solved
in deterministic polynomial time.  In the remaining cases, if $f\neq
g^d$, then there exists a leaf $u$ in $T$ labeled by a non-dual pair
$(\gvp',\psi')$. If $seq(u)$ is known, we can compute, by simulating
$A$ on the branch described by $seq(u)$, the entire path
$u_0,u_1,\ldots,u_l = u$ from the root $u_0$ to $u$ with all labels
$I(u_0) = (\gvp^*,\psi^*)$, $I(u_1)$, \ldots, $I(u_l)$ and check that
$I(u_l)$ is non-dual in steps 2 and 3 of A in polynomial time. Since
the binary length of any standard encoding of $(\gvp^*,\psi^*)$ is
polynomially related to $n=|\gvp^*|+|\psi^*|$ if $seq(u)$ is nonempty,
to prove the result it is sufficient to show that $seq(u)$ can be 
constructed in polynomial time from $O(\log^3 v^*)$ suitably guessed
bits.  To see this, let us represent every $seq(u)$ as a sequence
$seq^*(u)={\tt [}\ell_0,\ell_1,\ell_2\ldots, \ell_k{\tt ]}$, where
$\ell_0$ is the number of leading right moves and $\ell_i$ is the
number of consecutive right moves after the $i$-th left move in
$seq(u)$, for $i=1,\ldots,k$. For example, if $seq(u)={\tt
[r,r,l,r,r,r,l]}$, then $seq^*(u)={\tt [}2,3,0{\tt ]}$.  By
Lemma~\ref{lem:e}, $seq^*(u)$ has length at most $\log^2 v^*+1$.
Thus, $seq^*(u)$ occupies in binary only $O(\log^3v)$ bits; moreover,
$seq(u)$ is trivially computed from $seq^*(u)$ in polynomial time.
\qed

\nop{ *** SHOULD WE KEEP THIS REMARK ? WE HAVE A SHARPER ONE ***
\begin{remark}
\rm It also follows that if $f\neq g^d$, then a witness $w$
can be found in polynomial time within $O(\log^3 n)$ nondeterministic
steps. In fact, the sequence $seq(u)$ to a ``failing leaf'' labeled
$(\gvp',\psi')$ describes a choice of values for all variables in
$V(\gvp\land\psi)\setminus V(\gvp'\land\psi')$. By completing
it with values for $V(\gvp'\land\psi')$ that show non-duality of
$(\gvp',\psi')$, we obtain in polynomial time a vector $w$ such that
$f(w)\neq g^d(w)$.
\end{remark}
}

\subsection{Analysis of Algorithm B of Fredman and Khachiyan}
\label{subsec:b}

The aim of the above proof was to exhibit a new method of algorithm
analysis that allows us to show with very simple means that duality
can be polynomially checked with limited nondeterminism.  By applying
the same method of analysis to the slightly more involved algorithm~B
of~\cite{fred-khac-96} (which runs in $n^{4\chi(n)+O(1)}$ time, and
thus in $n^{o(\log n)}$ time), we can sharpen the above result by
proving that deciding whether monotone CNFs $\gvp$ and $\psi$ are
non-dual is feasible in polynomial time with $O(\chi(n)\cdot\log n)$
nondeterministic steps; consequently, the problem {\sc Dual} is in
$\cobetapol{2}$.

Like algorithm~A, also algorithm~B uses a recursive self-reduction
method that decomposes its input, a pair $(\gvp,\psi)$ of monotone
CNFs, into smaller inputs instances for recursive calls. Analogously,
the algorithm is thus best described via its {\em recursion tree} $T$,
whose root represents the input instance $(\gvp^*$, $\psi^*)$ (of size
$n$), whose intermediate nodes represent smaller instances, and whose
leaves represent those instances that can be solved in polynomial
time. Like for algorithm~A, the nodes $u$ in $T$ are labeled with the
respective instances $I(u)=(\gvp,\psi)$ of monotone pairs. Whenever
there is a branching from a node $u$ to children, then $I(u)$ is a
pair of dual monotone CNFs iff $I(u')$ for {\em each} child $u'$ of
$u$ in $T$ is a pair of dual monotone CNFs. Therefore, the original
input $(\gvp^*$, $\psi^*)$ is a dual monotone pair iff all leaves of
$T$ are labeled with dual monotone pairs.

Rather than describing algorithm~B in full detail, we confine here to
recall those features which are relevant for our analysis.  In
particular, we will describe some essential features of its recursion
tree $T$. 


For each variable $x_i$ occurring in $\gvp$, the {\em frequency $\efi$
of $x_i$ w.r.t. $\gvp$} is defined as $\efi= \frac{|\{c\in \gvp\,:\;
x_i\in c\}|}{|\gvp|}$, i.e., as the number of clauses of $\gvp$
containing $x_i$ divided by the total number of clauses in
$\gvp$. Moreover, for each $v\geq 1$, let $\chi(v)$
be defined by $\chi(v)^{\chi(v)}=v$.
%
%

Let $v^*=|\gvp^*||\psi^*|$ denote the volume of the input (=root)
instance $(\gvp^*,\psi^*)$. For the rest of this section, we assume
that $|\gvp^*|+|\psi^*|\leq |\gvp^*|\cdot|\psi^*|$.  In fact, in any instance
which violates this inequality, either $\gvp^*$ or
$\psi^*$ has at most one clause; in this case, {\sc Dual}
is trivially solvable in polynomial time.

Algorithm~B first constructs the root $r$ of $T$ and then recursively
expands the nodes of $T$.  For each node $u$ with label 
$I(u) = (\gvp,\psi)$, algorithm~B does the following.

The algorithm first performs a polynomial time computation, which we shall 
refer to as {\sc LCheck}$(\gvp,\psi)$ here, as follows. {\sc LCheck}$(\gvp,\psi)$ first eliminates 
all redundant (i.e., non-minimal) clauses from $\gvp$ and $\psi$ and 
then tests whether some of the following conditions is violated:
\begin{enumerate}
\item $V(\gvp)=V(\psi)$; 
\item  $\max_{c\in\gvp}|c|\leq|\psi|$ \ and \  $\max_{c\in\psi}|c|\leq|\gvp|$; 
\item $\min(\,|\gvp|,\,|\psi|\,) > 2$.
\end{enumerate} 
If {\sc LCheck}$(\gvp,\psi)=\mathit{true}$, then $u$ is a leaf of $T$ (i.e.,
not further expanded); whether $I(\gvp,\psi)$ is a dual monotone pair is then
decided by some procedure {\sc Test}$(\gvp,\psi)$ in polynomial time.  In case
{\sc Test}$(\gvp,\psi)$ returns $\mathit{false}$, the original input
$(\gvp^*$, $\psi^*)$ is not a dual monotone pair, and algorithm B
returns $\mathit{false}$. Moreover, in this case a counterexample $w$
to the duality of $\gvp^*$ and $\psi^*$ is computable in polynomial
time from the path leading from the root $r$ of $T$ to $u$.

If {\sc LCheck}$(\gvp,\psi)$ returns $\mathit{false}$, algorithm~B chooses in
polynomial time some appropriate variable $x_i$ such that $\efi>0$ and
$\egi>0$, and creates two or more children of $u$ by deterministically
choosing one of three alternative decomposition rules {\bf (i)}, {\bf
(ii)}, and {\bf (iii)}. Each rule
decomposes $I(u)=(\gvp,\psi)$ into smaller instances, whose respective
volumes are summarized as follows. Let, as for algorithm~A,
$\gvp_0 = \{c-\{x_i\} \mid x_i\in c,\, c \in \gvp\},$ $\gvp_1 = \{c
\mid x_i\notin c,\, c\in \gvp\},$ $\psi_0 = \{ c'-\{x_i\} \mid x_i\in
c',\, c'\in \psi\}$, and $\psi_1 = \{c' \mid x_i\notin c',\, c'\in
\psi\}$. Furthermore, define $\ev = 1/\chi(v)$, for any $v > 0$.
\begin{description}
\item{\bf Rule (i)} 
If $\efi \leq \evar{v(u)}$, then $I(u)$ is decomposed into:
\begin{description}
\item[\bf a)] one instance $(\gvp_1,\psi_0\land \psi_1)$
of volume $\leq (1-\efi)\cdot v(u)$;
\item[\bf b)] $|\psi_0|$ instances  $I_1,\ldots,I_{|\psi_0|}$ of volume $\leq \efi \cdot v(u)$ each.
Each such instance $I_j$ corresponds to one clause of $\psi_0$ and can
thus be identified as the $j$-th clause of $\psi_0$ with an index $j\leq |\psi_0|<n$ 
(recall that  $n$ denotes 
the size of the original input). 
\end{description}

\item{\bf Rule (ii)}
If $\efi > \evar{v(u)} \geq \egi$,  then $I(u)$ is decomposed into:
\begin{description}
\item[\bf a)] one instance $(\psi_1,\gvp_0\land \gvp_1)$
of volume $\leq (1-\egi)\cdot v(u)$;
\item[\bf b)] $|\gvp_0|$ instances $I_1,\ldots,I_{|\gvp_0|}$  of volume $\leq \egi \cdot v(u)$ each.
Each such instance $I_j$ corresponds to one clause of $\gvp_0$ and can
be identified by an index $j\leq |\gvp_0|<v^*$.
\end{description}

\item{\bf Rule (iii)}
If both $\efi > \evar{v(u)}$ and $\egi> \evar{v(u)}$, then $I$ is decomposed into:
\begin{description}
\item[$\bf c_0)$] one instance of volume $\leq (1-\efi)\cdot v(u)$, and
\item[$\bf c_1)$] one instance of volume $\leq (1-\egi)\cdot v(u)$.
\end{description}
\end{description}
Algorithm~B returns {\em true} iff {\sc Test}$(I(u))$ returns {\em true}
for each leaf $u$ of the recursion tree. This concludes the
description of algorithm~B.

For each node $u$ and child $u'$ of $u$ in $T$, we label the 
arc $(u,u')$ with the precise type of rule that was used 
to generate $u'$ from $u$. The possible labels are thus 
{\bf (i.a)},  {\bf (i.b)},  {\bf (ii.a)},  {\bf (ii.b)}, {\bf (iii.c$_0$)}, and  
{\bf (iii.c$_1$)}. We call {\bf (i.a)} and {\bf (ii.a)} {\em a-labels},
{\bf (i.b)} and {\bf (ii.b)} {\em b-labels},
and {\bf (iii.c$_0$)} and {\bf (iii.c$_1$)} {\em c-labels}.
Any arc with a $b$-label is in addition labeled with the index $j$ of
the respective instance $I_j$ in the decomposition, which we refer to as the {\em $j$-label} of the arc.

\begin{definition}
For any node $u$ of the tree $T$,  let $seq(u)$ 
denote the sequence of all edge-labels on the path 
from the root $r$ of $T$ to $u$.
\end{definition}

Clearly, if $seq(u)$ is known, then the entire path from $r$ to $u$
including all node-labels (in particular, the one of $u$) can be
computed in polynomial time.  Indeed, the depth of the tree is at most
$v^*$, and adding a child to a node of $T$ according to algorithm~B
is feasible in polynomial time.

The following lemma bounds the number of various labels which may
occur in $seq(u)$.

\begin{lemma}
\label{lem:e2}
For each node $u$ in $T$, $seq(u)$ contains at most (i) $v^*$ many
$a$-labels, (ii) $\log v^*$ many $b$-labels, and (iii)  $\log^2 v^*$ many $c$-labels.
\nop{******** old ********
\begin{enumerate}
\item[a)] The number of $a$-labels in $seq(u)$ 
is bounded by $v^*$.

\item[b)] The number of $b$-labels in $seq(u)$ 
is bounded by $\log v^*$.

\item[c)] The number of $c$-labels in $seq(u)$ is bounded by  $log^2 v^*$.
\end{enumerate}
********}
\end{lemma}

\begin{proof}
\noindent (i) Let us consider rule {\bf (i.a)} first.  Given that
$\efi>0$, $x_i$ effectively occurs in some clause of $\gvp$. Thus
$|\gvp_1|<|\gvp|$. Moreover, by definition of $\psi_0$ and $\psi_1$,
$|\psi_0\land \psi_1|\leq |\psi|$. Thus we have $|\gvp_1|\cdot|\psi_0\land
\psi_1|< |\gvp|\cdot|\psi|$. It follows that whenever rule {\bf (i.a)} is
applied, the volume decreases (at least by 1). The same holds for rule
{\bf (ii.a)} by a symmetric argument. Since no rule ever increases the
volume, there are at most $v^*$ applications of an $a$-rule.

\noindent (ii) Assume that rule {\bf (i.b)} is applied to generate a
child $t'$ of node $t$. By condition~3 of {\sc LCheck},
$v(t)>4$. Therefore, $\chi(v(t))>2$ and thus $\efi\leq \evar{v(t)}
<1/2$.  It follows that $v(t')<v(t)/2$. The same holds if $t'$ results
from $t$ via rule $\bf (ii.b)$.  Because no rule ever increases the
volume, any node generated after (among others) $\log v^*$
applications of a $b$-rule has volume $\leq$ 1 and is thus a leaf in $T$.

\noindent (iii) If a $c$-rule is applied to generate a child $t'$ of a node $t$,
and since $\evar{v(t)} > \evar{v^*} > 1/\log v^*$, the volume of $v(t)$
decreases at least by factor $(1- 1/\log v^*)$. Thus, the volume
of any node $u$ which results from $t$ after  $\log v^*$  applications of a
$c$-rule satisfies $v(u)\leq v(t)(1- 1/\log v^*)^{\log v^*}\leq v(t)/e$
by (\ref{eq-limes}); i.e., the volume has decreased more than
half. Thus, any node $u$ resulting from the root of $T$ after $\log^2 v^*$ applications of a $c$-rule satisfies $v(u)
\leq v^*\cdot \Big(\frac{1}{2}\Big)^{\log v^*} = 1$; that is, $u$ is a
leaf in $T$. 
\qed
\end{proof}

\begin{theorem}
\label{theo:beta2}
Deciding whether monotone CNFs $\gvp$ and $\psi$ are non-dual is
feasible in polynomial time with $O(\log^2 n)$ nondeterministic steps,
where $n=|\gvp|+|\psi|$.  
\end{theorem}

\begin{proof}
As in the proof of Theorem~\ref{theo:beta3}, we use a compact representation 
$seq^*(u)$ of $seq(u)$. However, here the definition of $seq^*$ is somewhat 
more involved: 
\begin{itemize}
 \item $seq^*(u)$ contains all $b$-labels of $seq(u)$, which are the anchor elements of $seq^*(u)$.
          Every $b$-label is immediately followed by its associated $j$-label, 
i.e., the label specifying which of the (many) $b$-children 
is chosen. We call a $b$-label and its associated $j$-label a {\em $bj$-block}. 

\item At the beginning of $seq^*(u)$, as well as 
after each $bj$-block, there is an {\em $ac$-block}. The first
 $ac$-block in $seq^*(u)$ represents the sequence of all $a$- and $c$-labels in $seq(u)$ preceding the 
first $b$-label in $seq(u)$, and the $i$-th $ac$-block in $seq^*(u)$,
$i> 1$,  represents the
sequence of the $a$ and $c$ labels (uninterrupted by any other label)
following the $(i-1)$-st $bj$-block in $seq(u)$. 

Each $ac$-block consists of an $\alpha$-block followed by a
        $\gamma$-block, where
\begin{itemize}
\item 
the $\alpha$-block contains, in binary, the {\em number} of $a$-labels in the 
$ac$-block, and   
\item the $\gamma$-block contains all $c$-labels (single bits) in the $ac$-block, in the order as they
appear.
\end{itemize}
\end{itemize}

For example, if $s$ = ``$(i.a),(ii.a),c_0,(ii.a),c_1,c_0,(i.a)$'' is a maximal
$ac$-subsequence in $seq(u)$, then its corresponding $ac$-block in $seq^*(u)$
is ``$10,c_0,c_1,c_0$'', where 10 (= 4) is the $\alpha$-block (stating that there 
are four $a$-labels) and 
``$c_0,c_1,c_0$'' is the $\gamma$-block enumerating the $c$-labels 
in $s$ in their correct order.

The following facts are now the key to the result. 

\begin{description}
\item[Fact~A.] Given $\phi^*$,$\psi^*$ and a string $s$, it is
possible to compute in polynomial time the path $r=u_0,u_1,\ldots,$
$u_l=u$ from the root $r$ of $T$ to the unique node $u$ in $T$ such
that $s = seq^*(u)$ and all labels $I(u_i)$, or to tell that no such
node $u$ exists (i.e., $s \neq seq^*(u)$ for every node $u$ in $T$).
\end{description}

This can be done by a simple procedure, which incrementally
constructs $u_0$, $u_1$, etc as follows. 

Create the root node $r=u_0$, and set $I(u_0)=(\phi^*,\psi^*)$ and
$t:=0$. Generate the next node $u_{t+1}$ and label it, while
processing the main blocks ($ac$-blocks and
$bj$-blocks) in $s$ in order, as follows: 

\begin{description}
\item[$ac$-block:] Suppose the $\alpha$-block of the current
$ac$-block has value $n_\alpha$, and the $\gc$-block contains labels
$\gc_1,\ldots,\gc_k$. Set up counters $p:=0$ and
$q:=0$, and while $p<n_\alpha$ or $q<k$, do the following.

If {\sc LCheck}$(I(u_t))=\mathit{true}$, then flag an error and halt, as $s\neq
seq^*(u)$ for every node $u$ in $T$. Otherwise,  
determine the rule type $\tau\in \{{\bf (i)},
{\bf  (ii)}, {\bf (iii)}\}$ used by algorithm~B to (deterministically) decompose $I(u_t)$. 
\begin{itemize}
\item If $\tau\in\{{\bf (i)},{\bf (ii)}\}$ and $p<n_\alpha$, then assign $I(u_{t+1})$ the $a$-child of $I(u_t)$ according
to algorithm~B, and increment $p$ and $t$ by 1.
\item If $\tau={\bf (iii)}$ and $q<k$, then increment $q$ by 1, assign
$I(u_{t+1})$ the $\gc_q$-child of $I(u_t)$ according
to algorithm~B, and increment $t$ by 1.
\item In all other cases (i.e.,  either $\tau\in\{{\bf (i)},{\bf (ii)}\}$ and $p\geq n_\alpha$,
or $\tau={\bf (iii)}$ and $q\geq k$), flag an error and halt, since $s\neq
seq^*(u)$ for every node $u$ in $T$.
\end{itemize}

\item[$bj$-block:]
Determine the rule type $\tau\in \{ {\bf (i)}, {\bf(ii)}, {\bf
(iii)}\}$ used by
algorithm~B to (deterministically) decompose $I(u_t)$. If
$\tau={\bf (iii)}$, then flag an error and halt,  since $s\neq seq^*(u)$ for
every node $u$ in $T$. Otherwise,  assign $I(u_{t+1})$ the $j'$-th
$\bf($$\tau.$$\bf b)$-child of $I(u_t)$ according to rule $\bf
($$\tau$$\bf.b)$ of
algorithm~B, where $j'$ is the $j$-label of the current $bj$-block. 
\end{description}

Clearly, this procedure outputs in polynomial time the desired labeled
path from $r$ to $u$, or flags an error if $s\neq seq^*(u)$ for every
node $u$ in $T$.

Let us now bound the size of $seq^*(u)$  in terms of the 
original input size $v^*$. 

\begin{description}
\item[Fact~B.] For any $u$ in $T$, the size of $seq^*(u)$ is $O(\log^2 v^*)$. 
\end{description}

By Lemma~\ref{lem:e2} (ii), there are $<\log v^*$ $bj$-blocks. As already
noted, each $bj$-block has size $O(\log v^*)$; thus, the total size of
all $bj$-blocks is $O(\log^2 v^*)$. Next, there are at most $\log v^*$ many
$ac$-blocks and thus $\alpha$-blocks. Each $\alpha$-block encodes a number of $< v^*$ $a$-rule applications (see
Lemma~\ref{lem:e2}.(i)), and thus uses at most $\log v^*$ bits.  The
total size of all $\alpha$-blocks is thus at most $\log^2
v^*$. Finally, by Lemma~\ref{lem:e2} (iii), the total size of all $\gamma$-blocks is at most $\log^2 v^*$.
Overall, this means that $seq^*(u)$ has size $O(\log^2 v^*)$.

To prove that algorithm~B rejects input $(\gvp^*,\psi^*)$, it is thus
sufficient to guess $seq^*(u)$ for some leaf $u$ in $T$, to compute in
polynomial time the corresponding path $r=u_0,u_1,\ldots,u_l=u$, and
to verify that {\sc LCheck}$(I(u))=\mathit{true}$ but {\sc
Test}$(I(u))=\mathit{false}$. Therefore, non-duality of $\phi^*$ and
$\psi^*$ can be decided in polynomial time with $O(\log^2 v^*)$ bit
guesses. Given that $v^*\leq n^2$, the number of guesses is $O(\log^2
n^2)=O(\log^2 n)$.  \qed
\end{proof}

The following result is an immediate consequence of this theorem.

\begin{corollary}
Problem {\sc Dual} is in $\cobetapol{2}$ and solvable in
deterministic $n^{O(\log n)}$ time,
where $n = |\gvp| + |\psi|$. 
%
%
\end{corollary}

(Note that Yes-instances of {\sc Dual} must have size polynomial in
$n$, since dual monotone pairs $(\gvp,\psi)$ must satisfy conditions (2)
and (3) in step~2 of algorithm~A.) We remark that the proof of Lemma~\ref{lem:e2} and
Theorem~\ref{theo:beta2} did no stress the fact that
$\evar{v}=1/\chi(v)$; the proofs go through for $\evar{v}=1/\log v$
as well. Thus, the use of the $\chi$-function is not
essential for deriving Theorem~\ref{theo:beta2}.

However, a tighter analysis of the size of $seq^*(u)$ stressing
$\chi(v)$ yields a better bound for the number of nondeterministic
steps.  In fact, we show in the next result that $O(\chi(n)\cdot\log
n)$ bit guesses are sufficient. Note that $\chi(n) = o(\log n)$,
thus the result is an effective improvement. Moreover, it also shows
that {\sc Dual} is most likely not complete for $\cobetapol{2}$.

\begin{theorem}
\label{theo:logchi}
Deciding whether monotone CNFs $\gvp$ and $\psi$ are non-dual is
feasible in polynomial time with $O(\chi(n)\log n)$ nondeterministic
steps, where $n = |\gvp|+|\psi|$.
\end{theorem}

\begin{proof}
In the proof of Theorem~\ref{theo:beta2}, our estimates of the
components of $seq^*(u)$ were rather crude. With more effort, we
establish the following.

\begin{description}
\item[Fact~C.] For any $u$ in $T$, the size of $seq^*(u)$ is  $O(\chi(v^*)\cdot\log(v^*))$.
\end{description}

Assume node $u'$ in $T$ is a child of $u$ generated via a
$b$-rule. The $j$-label of the arc $(u,u')$ serves to identify one
clause of $I(u)$. Clearly, there are no more than $v(u)$
such clauses. Thus $\log v(u)$ bits suffice to represent any $j$-label.

Observe that if $u$ is a node of $T$, then any path $\pi$ from $u$ to
a node $w$ in $T$ contains at most $v(u)$ nodes, since the volume
always decreases by at least 1 in each decomposition step.  Thus, the
number of $a$-labeled arcs in $\pi$ is bounded by $v(u)$ and not just
by $v^*$ ($=v(r)$).

For each node $u$ and descendant $w$ of $u$ in $T$, let
$$
f(u,w) = \sum_{u'\in B(u,w)} \log v(u'),
$$
where $B(u,w)$ is the set of all nodes $t$ on the path from $u$ 
to $w$ such that the arc from $t$ to its successor on the path is $b$-labeled.

By what we have observed, the total size of all encodings of
$j$-labels in $seq^*(u)$ is at most $f(v^*,u)$ and the size of all
$\alpha$-blocks in $seq^*(u)$ is at most $\log(v^*) + f(v^*,u)$, were
the first term takes care of the first $\alpha$-block and the second
of all other $\alpha$-blocks. Therefore, the total size of all
$\alpha$-blocks and all $bj$-blocks in $seq^*(u)$ is
$O(f(v^*,u)+\log(v^*))$.

We now show that for each node $u$ and descendant $w$ of $u$ in $T$, it holds that
$$
f(u,w)\leq \log(v(u))\cdot \chi(v(u)).
$$
The proof is by induction on the number $|B(u,w)|$ of $b$-labeled arcs
on the path $\pi$ from $u$ to $w$.  If $|B(u,w)|=0$, then obviously
$f(u,w)=0\leq v(u)$.

Assume the claim holds for $|B(u',w)|\leq i$ and consider
$|B(u,w)|=i+1$.  Let $t$ be the first node on $\pi$ contained in
$B(u,w)$, and let $t'$ be its child on $\pi$.  Clearly,
$f(u,w)=f(t,w)$, and thus we obtain:
\begin{tabbing}
X\= $f(u,w)$  \= $\leq$ \= $\log(v(t)) + (\log(v(t)-\log(\chi(v(t)))\cdot\chi(v(t))$ \quad \= \kill 
\> $f(u,w)$ \> = \> $\log(v(t)) + f(t',w)$ \+\+\\[0.75ex] 
 $\leq$ \> $\log(v(t)) + \log(v(t'))\cdot\chi(v(t'))$ \> (\textrm{induction hypothesis}) \\[0.75ex] 
 $\leq$ \> $\log(v(t)) + (\log(v(t))-\log(\chi(v(t))))\cdot\chi(v(t))$ \>
       (\textrm{as $v(t') \leq \frac{v(t)}{\chi(v(t))}$,\ $\chi(v(t'))\leq \chi(v(t))\,$})\\[0.75ex] 
 = \> $\log(v(t))\cdot \chi(v(t))$ \>  (\textrm{as
$\log(\chi(y))\cdot\chi(y) = \log y$, for all $y$}). 
\end{tabbing}
Thus, $f(u,w) \leq \log(v(u))\cdot \chi(v(u))$. This
concludes the induction and proves the claim.

Finally, we show that the total size of all $\gamma$ blocks in
$seq^*(u)$, i.e., the number of all $c$-labels in $seq(u)$, is bounded
by $\chi(v^*)\cdot\log(v^*)<\log^2 v^*$. Indeed, assume a $c$-rule is
applied to generate a child $t'$ of any node $t$, and let $v=v(t)$, $v'=v(t')$. Since $\efi >
\evar{v}$ and $\egi>\evar{v}$, we have
$v'<(1-\evar{v})\cdot v$. Since $\chi(v^*)>\chi(v)$, we have
$\evar{v} = 1/\chi(v) > 1/\chi(v^*)$ and thus
$$
v'<\Big(1-\frac{1}{\chi(v^*)}\Big)\cdot v.
$$ 
Hence, any node in $T$ 
resulting after $\chi(v^*)\cdot\log(v^*)$ applications
of a $c$-rule has volume at most
$$
v^*\cdot\Big(1-\frac{1}{\chi(v^*)}\Big)^{\chi(v^*)\cdot\log v^*}
 = v^*\cdot\Big[\Big(1-\frac{1}{\chi(v^*)}\Big)^{\chi(v^*)}\Big]^{\log v^*}
 \leq v^*\cdot\Big(\frac{1}{e}\Big)^{\log v^*}\;\leq\; 1
$$
(cf.\ also (\ref{eq-limes})). Consequently, along each branch in $T$
there must be 
no more than $\chi(v^*)\cdot\log v^*$ applications of a $c$-rule.  In summary, the
total sizes of all $\alpha$-blocks, all $\gamma$-blocks, and  all
encodings of $j$-labels in $seq^*(u)$ are all bounded by
$\chi(v^*)\cdot\log v^*$. This proves Fact~C.

As a consequence, non-duality of a monotone pair $(\gvp^*,\psi^*)$ can be recognized in
polynomial time with $O(\chi(v^*)\cdot\log v^*)$ many bit guesses. As
already observed on the last lines of~\cite{fred-khac-96}, we have
$\chi(v^*)<2\chi(n)$. Furthermore, $v^*\leq n^2$, thus $\log v^*\leq
2\log n$. Hence, non-duality $(\gvp^*,\psi^*)$ can be recognized in polynomial time with
$O(\chi(n)\cdot\log(n))$ bit guesses.  \qed
\end{proof}

\begin{corollary}
Problem {\sc Dual} is solvable in deterministic $n^{O(\chi(n))}$ time,
where $n = |\gvp| + |\psi|$. 
%
%
\end{corollary}

\begin{remark}\rm 
Note that the sequence $seq(u)$ describing a path from the root of $T$
to a ``failure leaf'' with label $I(u)=(\gvp',\psi')$
describes a choice of values for all variables in
$V(\gvp\land\psi)\setminus V(\gvp'\land\psi')$. By completing it with
values for $V(\gvp'\land\psi')$ that show non-duality of
$(\gvp',\psi')$, which is possible in polynomial time, we obtain in
polynomial time from $seq(u)$ a vector $w$ such that $f(w)\neq
g^d(w)$. It also follows from the proof of Theorem~\ref{theo:logchi} that a witness $w$ for $f\neq g^d$ (if
one exists) can be found in polynomial time with $O(\chi(n)\cdot\log n)$
nondeterministic steps. 
\end{remark}
  
\subsection{Application of Beigel and Fu's results}
\label{sec:beigel-fu}

While our independently developed methods substantially differ from
those in \cite{beig-fu-97,beig-fu-99}, membership of problem {\sc
Dual} in $\cobetapol{2}$ may also be obtained by exploiting Beigel and
Fu's Theorem~8 in~\cite{beig-fu-97} (or, equivalently, Theorem~11
in~\cite{beig-fu-99}). They show how to convert certain recursive
algorithms that use disjunctive self-reductions, have runtime bounded
by $f(n)$, and fulfill certain additional conditions, into polynomial
algorithms using $\log(f(n))$ nondeterministic steps (cf.\
\cite[Section 5]{beig-fu-99}). 


Let us first introduce the main relevant definitions of
\cite{beig-fu-97}. Let $\|y\|$ denote the size of a problem instance
$y$.

\begin{definition}[\cite{beig-fu-97}]
\label{defn:wellf}
A partial order $\prec$ (on problem instances) 
is polynomially well-founded, if there exists a polynomial-bounded 
function $p$ such that 
\begin{itemize}
\item $y_m\prec\cdots\prec y_1 \Rightarrow m\leq p(\|y_1\|)$ and
\item $y_m\prec\cdots\prec y_1 \Rightarrow \|y_m\|\leq  p(\|y_1\|)$.
\end{itemize}
\end{definition}
For technical simplicity, \cite{beig-fu-97} considers only languages
(of problem instances) containing the empty string, $\Lambda$.

\begin{definition}[\cite{beig-fu-97}] 
\label{defn:dsr}
A disjunctive self-reduction (for short, d-self-reduction)  
for a language $L$ is a pair $\tuple{h,\prec}$ of a polynomial-time 
computable function $h(x)=\{x_1,\ldots,x_m\}$ and a 
polynomially well-founded partial order $\prec$ on problem instances
such that 
\begin{itemize}
\item $\Lambda$ is the only minimal element under $\prec$;
\item for all $x\neq\Lambda$, $x\in L\equiv h(x)\cap L\neq\emptyset$;
\item for all $x$, $x_i\in h(x)\Rightarrow x_i\prec x$.
\end{itemize}
\end{definition}

\newcommand{\rect}{{{\rm REC}(T(x))}} 
\newcommand{\thp}[1]{{T_{h,\prec}(#1)}}

\begin{definition}[\cite{beig-fu-97}] 
Let $\tuple{h,\prec}$ be a d-self-reduction and let $x$ be a problem instance.
\begin{itemize}
\item $\thp{x}$ is the unordered rooted tree that satisfies the
following rules: (1) the root is $x$; (2) for each $y$, the set of
children of $y$ is $h(y)$.
\item $|\thp{x}|$ is the number of leaves in 
$T_{h,\prec}(x)$.
\end{itemize}
\end{definition}

\begin{definition}[~\cite{beig-fu-97}] 
\label{defn:rect}
Let $T$ be a polynomial-time computable function. A language $L$ is in
$\rect$, if there is a d-self-reduction $\tuple{h,\prec}$ for
$L$ such that for all $x$
\begin{enumerate}
\item $|\thp{x}|\leq T(x)$, and
\item $T(x)\geq \sum_{x_i\in h(x)}T(x_i)$.
\end{enumerate}
\end{definition}

Let $\nondet{T(x)}$ denote the set of all (languages of) problems
whose Yes-instances $x$ are recognizable in  polynomial time 
with $T(x)$ nondeterministic bit guesses.

\begin{theorem}[~\cite{beig-fu-97}] 
\label{theo:beigelfu}
$\rect\subseteq\nondet{\lceil\log T(x)\rceil}$ 
\end{theorem}

We now show that Theorem~\ref{theo:beigelfu}, together with 
Fredman's and Khachiyan's proof of the deterministic complexity 
of algorithm~B, can be used to prove that problem {\sc Dual} is in 
co-$\betapol{2}$.

Let $L$ denote the set of all non-dual monotone pairs $(\gvp,\psi)$
plus $\Lambda$. Let us identify each monotone pair $(\gvp,\psi)$ which
satisfies {\sc LCheck}$(\gvp,\psi)$ but does not satisfy {\sc
Test}$(\gvp,\psi)$ with the ``bottom element'' $\Lambda$. Thus, if a
node in the recursion tree $T$ has a child labeled with such a pair, then
the label is simply replaced by $\Lambda$.

Let us define the order $\prec$ on monotone pairs plus $\Lambda$ as
follows: $J\prec I$, if $I\neq J$ and either $J=\Lambda$ or $J$ labels
a node of the recursion tree generated by algorithm~B on input $I$. It
is easy to see that both conditions of Definition~\ref{defn:wellf}
apply; therefore, $\prec$ is polynomially well-founded. In fact, we
may define the polynomial $p$ by the identity function; since the
sizes of the instances in the recursion tree strictly decrease on
each path in $T$, the two conditions hold.

Define $h$ as the function which associates 
with each monotone pair $I=(\gvp,\psi)$ 
those instances that label all children of the root 
by algorithm $B$ on input $I$. Clearly $h$ satisfies all three
conditions of Definition~\ref{defn:dsr}, and hence $\tuple{h,\prec}$ is a 
$d$-self-reduction for $L$.

Let $T$ be the function which to each instance $I$ associates
$v(I)^{\log v(I)}$ (recall that $v(I)$ denotes the volume of
$I$). It is now sufficient to check that conditions 1 and 2 of
Definition~\ref{defn:rect} are satisfied, and to ensure that
Theorem~\ref{theo:beigelfu} can be applied.

That item~1 of Definition~\ref{defn:rect} is satisfied follows
immediately from Lemma~5 in~\cite{fred-khac-96}, which states that the
maximum number of recursive calls of algorithm~B on any input $I$ of
volume $v$ is bounded by $v^{\chi(v)}$ $( \leq v^{\log v})$.  Retain,
however, that the proof of this lemma is noticeably more involved than
our proof of the membership of {\sc Dual} in co-$\betapol{2}$.

To verify item~2 of Definition~\ref{defn:rect}, it is sufficient
to prove that for a volume $v > 4$ of any input instance 
to algorithm $B$, it holds that
\begin{eqnarray}
v^{\log v}  &\geq & (v-1)^{\log (v-1)} + 
\frac{v}{3}\cdot\Big(\frac{v}{2}\Big)^{\log \frac{v}{2}},\quad \textrm{ and} \label{i+ii} \\ 
v^{\log v}  &\geq & 2(\alpha\cdot v)^{\log (\alpha\cdot v)},\quad
\textrm{ where $\alpha = 1-1/\log v$}; \label{iii}
\end{eqnarray}
here, (\ref{i+ii}) arises from the rules ${\bf (i)}$, ${\bf (ii)}$ and
(\ref{iii}) from rule ${\bf (iii)}$. As for (\ref{i+ii}), the
child of $u$ from ${\bf (i.a)}$ resp.\ ${\bf (ii.a)}$ has volume at most $v-1$, and there are at most
$v/3$ many children from${\bf (i.b)}$ resp.\ ${\bf (ii.b)}$, since $\min(|\gvp|,|\psi|)>2$ (recall
that $v = |\gvp|\cdot|\psi|$);
furthermore, each such child has volume $\leq \ev\cdot v \leq
\frac{1}{2}v$.  In case of (\ref{iii}), the volume of each child of
$u$ is bounded by $(1-\ev)\cdot v \leq (1-1/\log v)\cdot v$; note also
that $v^{\log v}$ monotonically increases for $v>4$.  To see
(\ref{i+ii}), we have 
\begin{tabbing}
XX\= $(v-1)^{\log(v-1)}+\frac{v}{3}\cdot\Big(\frac{v}{2}\Big)^{\log \frac{v}{2}}$  \= $\leq$ \= $2(\frac{1}{e}\cdot \alpha^{\log v + \log \alpha})\cdot v^{\log v + \log \alpha}$ \quad \= \kill 
\> $(v-1)^{\log(v-1)}+\frac{v}{3}\cdot\Big(\frac{v}{2}\Big)^{\log
    \frac{v}{2}}$ \> $\leq$ \> $(v-1)^{\log v} + \frac{v}{3}\cdot\frac{v^{\log v-1}}{2^{\log v-1}}$ \+ \\[0.75ex]
    \> = \>  $v^{\log v}\cdot(1-\frac{1}{v})^{\log v} + \frac{2\cdot v^{\log v}}{3\cdot v}$ \\[0.75ex]
    \> $\leq$ \> $v^{\log v}\cdot(1-\frac{1}{v}+\frac{2}{3\cdot v})$ \\[0.75ex]
    \> = \> $v^{\log v}\cdot(1- \frac{1}{3\cdot v})$ \\[0.75ex]
    \> $<$ \>  $v^{\log v}$;\-
\end{tabbing}
to show (\ref{iii}), note that
\begin{tabbing}
XX\= $2(\alpha\cdot v)^{\log(\alpha\cdot v)}$  \= $\leq$ \= $2(\frac{1}{e}\cdot \alpha^{\log \alpha})\cdot
    v^{\log v + \log \alpha}$ \quad \= \kill 
\> $2(\alpha\cdot v)^{\log(\alpha\cdot v)}$ \> = \>  $2\alpha^{\log v + \log \alpha}\cdot v^{\log v + \log \alpha}$ \+ \\[0.75ex]
    \> $\leq$ \>  $2(\frac{1}{e}\cdot \alpha^{\log \alpha})\cdot
    v^{\log v + \log \alpha}$  \> ($\alpha^{\log v}\leq 1/e$,~ by (\ref{eq-limes}))\\[0.75ex]
    \> = \> $\frac{2}{e}\cdot (\alpha\cdot v)^{\log \alpha}\cdot v^{\log
    v}$ \\[0.75ex]
    \> $\leq$ \> $\frac{2}{e}\cdot v^{\log v}$ \> $(\alpha\cdot
    v)^{\log \alpha}\leq 1$,~ i.e.,~ $\log\alpha\cdot(\log\alpha+\log v)\leq 0$,\\
    \>\>\> since ${-1}<\log \alpha \leq 0$ and $\log v > 2$ \\[0.75ex]  
    \> $<$ \>  $v^{\log v}$. \-
\end{tabbing}
We can thus apply Theorem~\ref{theo:beigelfu} and conclude 
that the complement of {\sc Dual} 
is in $\nondet{\lceil\log T(x)\rceil}$, and thus also 
in $\betapol{2}$.

The advantage of Beigel and Fu's method is its very abstract 
formulation. The method has two disadvantages, however, that are 
related to the two items of Definition~\ref{defn:rect}.

The first item requires that $T(x)$ is at least the number of
leaves in the tree for $x$. In order to show this, one must basically
prove a deterministic time bound for the considered algorithm (or at
least a bound of the number of recursive calls for each instance,
which is often tantamount to a time-bound).  The method does not
suggest how to do this, but presupposes that such a bound exists (in
the present case, this was done by Fredman and Khachiyan in a
nontrivial proof).  The second item requires to prove that the
$T$-value of any node $x$ in the recursion tree is at least the sum of
the $T$-values of its children.  This may be hard to show in many cases, and does not necessarily hold for every upper bound $T$.

Our method instead  does not require an a priori time bound, but 
directly constructs a nondeterministic algorithm from 
the original deterministic algorithm, which lends itself to 
a simple analysis that directly leads to the desired 
nondeterministic time bound. The deterministic time bound 
follows as an immediate corollary. It turns out (as exemplified 
by the very simple proof of Theorem~\ref{theo:beta2}) that the
analysis involved in our method can be simpler than 
an analysis according to previous techniques.

\section{Conclusion}

We have presented several new cases of the monotone dualization
problem which are solvable in output-polynomial time. These cases
generalize some previously known output-polynomial
cases. Furthermore, we have shown by rather simple means that
non-dual monotone pairs $(\gvp,\psi)$ can be recognized, using a
nondeterministic variant of Fredman and Khachiyan's algorithm~B
\cite{fred-khac-96}, in polynomial time with $O(\log^2 n)$ many bit
guesses, which places problem {\sc Dual} in the class
$\cobetapol{2}$. In fact, a refined analysis revealed that this is
feasible in polynomial time with $O(\chi(n)\cdot \log n)$ many bit
guesses.

While our results document progress on {\sc Dual} and {\sc
Dualization} and reveal novel properties of these problems, the
question whether dualization of monotone pairs $(\gvp,\psi)$ is
feasible in polynomial time remains open. It would be interesting to
see whether the amount of guessed bits can be further significally
decreased, e.g., to $O(\log\log v\cdot \log v)$ many bits.

\subsection*{Acknowledgments}

Georg Gottlob thanks Jos\'e Balc\'azar for inviting him to an
interesting workshop in Bellaterra, Barcelona, Spain, and to Leonard
Pitt for giving an elucidating lecture on the dualization problem
there. This work was supported in part by the Austrian Science Fund
(FWF) project Z29-INF, by TU Wien through a scientific collaboration
grant, and by the Scientific Grant in Aid of the Ministry of
Education, Science, Sports and Culture of Japan.

{\small

\bibliographystyle{abbrv}

\ifmakebbl

\bibliography{library,eiterlib+}

\else

\fi

}

\end{document}